%% file: Risk_Verification_of_Stochastic_Systems_with_Neural_Network_Controllers.tex
\documentclass[11pt]{article}

\usepackage{lineno,hyperref}
\modulolinenumbers[5]

\usepackage{amsmath}
\usepackage{amssymb}
\usepackage{amsthm}
\usepackage{tikz}
\usepackage{pgfplots}
\usepackage{graphicx}
\usepackage{slashbox}
\usepackage{caption}
\usepackage{subcaption}
\usepackage{xcolor}
\usepackage{balance}
\allowdisplaybreaks  
\usepackage{cite}
\usepackage{algorithmicx}
\usepackage{algpseudocode}
\usepackage{algorithm}
\usepackage{subcaption}
\usepackage{hyperref}
\usepackage{comment}
\usepackage{tabularx}
\usepackage{bbm}
\usepackage{thmtools}

\allowdisplaybreaks

\renewcommand{\thmcontinues}[1]{cont.}

\newcommand{\specialcell}[2][c]{%
	\begin{tabular}[#1]{@{}c@{}}#2\end{tabular}}

\usepackage{array}
\usepackage{authblk}
\usepackage{setspace}
\usepackage{fullpage,etoolbox}

\newtheorem{example}{Example}

\input{commands.tex}

\title{Risk Verification of Stochastic Systems\\ with Neural Network Controllers\thanks{This research was generously supported by AFRL and DARPA FA8750-18-C-0090, AFOSR grant FA9550-19-1-0265, and NSF award CPS-2038873.} }
\author{Matthew Cleaveland$^\dagger$}
\author{Lars Lindemann$^\dagger$}
\author{Radoslav Ivanov$^+$}
\author{George J. Pappas$^\dagger$}
\affil{$^\dagger$ Department of Electrical and Systems Engineering, University of Pennsylvania}
\affil{$^+$ Department of Computer Science, Rensselaer Polytechnic Institute}

\begin{document}

\maketitle

\begin{abstract}
Motivated by the fragility of neural network (NN) controllers in safety-critical applications, we present a data-driven framework for verifying the risk of stochastic dynamical systems with NN controllers. Given a stochastic control system, an NN controller, and a specification equipped with a notion of trace robustness (e.g., constraint functions or signal temporal logic), we collect trajectories from the system that may or may not satisfy the specification. In particular, each of the trajectories produces a robustness value that indicates how well (severely) the specification is satisfied (violated). We then compute risk metrics over these robustness values to estimate the risk that the NN controller will not satisfy the specification. We are further interested in quantifying the difference in risk between two systems, and we show how the risk estimated from a nominal system can provide an upper bound for the risk of a perturbed version of the system. In particular, the tightness of this bound depends on the closeness of the systems in terms of the closeness of their system trajectories. For Lipschitz continuous and incrementally input-to-state stable systems, we show how to exactly quantify system closeness with varying degrees of conservatism, while we estimate system closeness for more general systems from data in our experiments. We demonstrate our risk verification approach on two case studies, an underwater vehicle and an F1/10 autonomous car.
\end{abstract}

\input{chapter/introduction}

\input{chapter/literature_review}

\input{chapter/problem_statement_new_3_30_2022} 

\input{chapter/risk_background}

\input{chapter/risk_gap}

\input{chapter/STL}

\input{chapter/case_studies}

\input{chapter/conclusion}

\section*{Acknowledgements}
This research was generously supported by AFRL and DARPA FA8750-18-C-0090, AFOSR grant FA9550-19-1-0265, and NSF award CPS-2038873.

\bibliographystyle{IEEEtran}
\bibliography{literature}

\appendix

\input{chapter/appendix}

\addtolength{\textheight}{-12cm}   

\end{document}

%% file: commands.tex
\newtheorem{theorem}{Theorem}
\newtheorem{lemma}{Lemma}

\newtheorem{proposition}{Proposition}
\newtheorem{corollary}{Corollary} 
\newtheorem{remark}{Remark}
\newtheorem{definition}{Definition} 
\newtheorem{problem}{Problem} 

\newcommand{\vertiii}[1]{{\vert\kern-0.25ex\vert\kern-0.25ex\vert #1\vert\kern-0.25ex\vert\kern-0.25ex\vert}}

\newtoggle{draft}
\togglefalse{draft}



%% file: chapter/introduction.tex
\section{Introduction}
\label{sec:introduction}



As autonomous systems become increasingly complex, formally reasoning about system properties becomes more challenging. In particular, verifying systems with machine learning (ML) components (e.g., for perception and control) is difficult as these are often synthesized without any formal guarantees. It has also been shown, see e.g., \cite{szegedy2013intriguing}, that ML components periodically fail which is especially concerning as these are increasingly being used in safety critical applications. Given this motivation, the goal of this work is: 1) to estimate the risk of stochastic systems with NN controllers in the loop from data, and 2) to provide conditions under which an NN controller has an acceptable risk for systems that are different from the data-generating system, e.g., due to changing environments that the system operates in. Risk metrics here allow for a trade off between the average and the worst possible case and can hence be used as a tuning knob towards risk-aware autonomy.

Depending on the application, there are several ways in which safety can be expressed. The simplest way is to express system safety as a constraint $c(x(t),t)\ge 0$  that has to hold for all $t\in\mathbb{T}$ for a time domain $\mathbb{T}\subseteq\mathbb{N}$, a constraint function $c:\mathbb{R}^n\times \mathbb{T}\to\mathbb{R}$, and a system trajectory $x: \mathbb{T} \to \mathbb{R}^n$. Importantly, one can calculate a robustness value $\inf_{t\in\mathbb{T}} c(x(t),t)$ for the trajectory $x(t)$ that serves as a robustness measure. When building autonomous or robotic systems, one is often interested in having the system perform more complex tasks and satisfy spatial and temporal properties that cannot be expressed by a constraint $c(x(t),t)\ge 0$, e.g., a service robot performing a series of re-active tasks with deadlines, while always remaining within a safe zone. A popular way to specify such behaviors is with temporal logics such as the commonly used linear temporal logic (LTL) \cite{pnueli1977temporal}. In LTL, a system trajectory either satisfies or violates an LTL specification, i.e., LTL has no notion of robustness. Recent effort has gone into adding a notion of robustness to LTL model checking \cite{Anevlavis2021} which, however, is quite coarse due to the discrete nature of LTL. As already argued above, it is important to quantify how robustly a trajectory satisfies a given specification. In the aforementioned  example, one is interested in distinguishing between trajectories in which the robot remains comfortably within the safe region versus those in which the robot skirts the edge of the safe region. In contrast to LTL, signal temporal logic (STL) allows one to assign a robustness value to a trajectory indicating how robustly it satisfies the specification \cite{maler2004monitoring}. Trajectories that robustly satisfy the specification are much less likely to violate it when perturbations or noise affect the system. This notion of STL robustness has been extended to stochastic systems in \cite{BARTOCCI2015} by considering the probability of satisfying the specification robustly. In our recent work \cite{lindemann2021stl}, we instead propose to use risk metrics towards considering the risk of not satisfying the specification robustly. This in particular allows for a trade-off between average and worst case scenarios.

\textbf{Contributions.} In this paper, we consider the setup in which we are given a stochastic dynamical control system, an NN controller, and a system specification that the NN controller aims to satisfy. The specification is expressed either as a constraint $c(x(t),t)\ge 0$ for all $t\in\mathbb{T}$ or as an STL specification. An example is an autonomous robot that is controlled by an NN  towards safely navigating an unknown environment. Our contributions are as follows:
\begin{enumerate}
    \item We formulate, for the first time, the risk verification problem of a closed-loop stochastic system with an NN controller by leveraging risk metrics and robustness values associated with the specification at hand.
    \item We estimate, from data, the risk of an NN controller not satisfying the specification robustly. The estimated risk has a clear interpretation that may be used to redesign the  controller in the case that the verification returns an unacceptably small (or even negative) robustness.
    \item We show how the risk estimated from a nominal system can provide an upper bound for the risk of a perturbed version of the system. We refer to the potential increase in risk as the risk verification gap. Particularly, the tightness of this risk verification gap depends on the closeness of the systems in terms of the closeness of their system trajectories. For Lipschitz continuous and incrementally input-to-state stable systems, we show how to exactly quantify system closeness with varying degrees of conservatism, while we estimate system closeness for more general systems from data in our experiments.
    \item We present  empirical evaluations for end-to-end perception-based reinforcement learning (RL) algorithms. We consider underwater vehicles as well as an autonomous racing car. We show how our risk verification approach can help to select the least risky controller from a set of controllers and empirically illustrate the risk verification gap as well as the effect of different risk metrics.
\end{enumerate}

We claim that our framework is practical in the following ways: 1) we are able to verify high-dimensional systems, 2) we directly verify the closed-loop system of an NN controlled stochastic system, and 3) we are able to quantify the risk verification gap when transferring from a nominal to a perturbed system.
We remark that NN controllers are complex and their closed-loop verification imposes significant computational costs, whereas our data-driven framework trades off this computational complexity with the ability to collect data. This framework differs from our previous work \cite{ivanov2020} because here we are verifying stochastic systems using a data-driven framework, whereas in \cite{ivanov2020} we had verified deterministic systems using a model-based approach.

The remainder of the paper is organized as follows. A literature review is given in Section \ref{sec:lit_review}. In Section \ref{sec:problem_statement}, we present the two systems considered in this paper (an F1/10 car and an unmanned underwater vehicle) and define the risk verification problem of these systems under NN controllers for constraint functions $c(x(t),t)$. In Section \ref{sec:sol_1}, we discuss different risk formulations and show how the risk of an NN controller can be estimated from data. Section \ref{sec:risk_gap} defines the risk verification gap between a nominal and a perturbed system and shows under which conditions the risk verification gap is small.  In Section \ref{sec:STL}, we show that our previous results extend in a similar manner when STL specifications are considered instead of constraint functions. We provide our two risk verification case studies in Section \ref{sec:caseStudies} and conclude the paper in Section \ref{sec:conclusion}.

%% file: chapter/literature_review.tex
\section{Literature Review}
\label{sec:lit_review}

Recently, statistical model checking (SMC) has been used to formally reason about complex systems and specifications in a computationally tractable way \cite{legay2019}. The general idea behind SMC is that one can use tools developed from the field of statistics, see e.g., \cite{Wald1945}, to analyze system traces and obtain guarantees with respect to the overall system behavior. SMC is popular because it is computationally lightweight, especially when compared to computationally expensive model checking and formal verification. This technique has been used to verify cyber-physical systems (CPS) under linear temporal logic (LTL) specifications \cite{Wang2019}, biological systems \cite{David2015}, and rare-event properties \cite{Jgourel2013}. A major advantage of the SMC regime is that its guarantees are generally based on finite data by providing probably approximately correct (PAC) guarantees. In other words, verification guarantees are such that with high confidence a given system satisfies a specification with a user-defined probability. Note  the appearance of two probabilities where particularly the first stems from the finite data regime, i.e., more data results in a  higher confidence.  


\textbf{Verification and Control under Risk.}
One drawback of SMC is that one can fall into the trap of only looking at the average performance of a system, whereas reasoning about system safety may require looking at the worst case performance (i.e. the tail end of the  distribution). Risk metrics, see e.g., \cite{Majumdar2020} for an introduction, offer a principled way of reasoning about the tail behaviors of a system and provide a meaningful tuning knob in between the two extremes. Risk metrics have been used to synthesize controllers \cite{wang2022risk} for the common linear quadratic regulator \cite{Tsiamis2020},  Markov decision processes (MDPs) \cite{chow2015risk}, constrained MDPs \cite{wachi2020safe} service robots in hospital settings \cite{Novin2020}, optimal energy grid management \cite{Khodabakhsh2016}, and reinforcement learning \cite{chow2017}. In addition, risk metrics are increasingly being used to reason about system safety and verification \cite{chapman2019risk,Chapman2021}. As an alternative to risk metrics, previous works have directly bounded the probability of a system violating a constraint, which are known as chance constraints. They have been used for MDPs \cite{ono2015chance}, partially observable Markov decision processes (POMDPs) \cite{thiebaux2016rao} and RL \cite{peng2021separated}. However, these approaches do not account for how robustly a controller satisfies its constraints as we do in this paper.

\textbf{Verification of ML components.} There is a rich literature on the verification of ML components. Most work falls into the open-loop verification category, which seeks to analyze the ML component on its own. For analyzing deep neural networks, most works focus on checking some form of input-output reachability \cite{Katz2017,Gehr2018, Dutta2018, Fazlyab2020, Wang2019,xue2022parametric}. In addition, work has gone into estimating the Lipschitz constants of deep neural networks \cite{fazlyab2019efficient}. Closed-loop verification, on the other hand, considers a dynamical system that is controlled by an NN controller in a feedback loop \cite{Katz2017}. For NN controllers, techniques for the closed-loop verification have appeared in \cite{Ivanov2019, Dutta2019,Fan2020,Tran2020,Huang2021}. Such formal reasoning approaches often assume full system knowledge, not allowing for stochastic components, and are conservative and computationally expensive. 

There also exist works using statistical methods to compute formal guarantees for open-loop ML components. For example, the authors in \cite{Vengertsev2020} use Monte-Carlo methods for verifying recurrent neural networks (RNNs). In \cite{Park2020}, confidence sets for neural networks are computed with PAC bounds and in \cite{Bates2021} an expected-loss minimization framework is used to create PAC bounded classifiers. In addition, work has gone into using statistical methods for estimating the robustness of neural networks \cite{webb2019, Mangal2019}. 

\textbf{Verification under Temporal Logic Specifications.} There exists a rich and mature body of literature on model checking under temporal logic specifications \cite{baier}. In the context of this work, we are, however, dealing with statistical model checking of temporal properties. In \cite{Zarei2020} and \cite{qin2022statistical}, the authors used statistical testing to derive high confidence bounds on the probability of a CPS with ML components satisfying an STL specification. In \cite{Wang2019}, LTL hyper-properties of CPS were verified using statistical techniques. Statistical techniques have also been used to verify discrete abstractions of hybrid systems \cite{Wang2015}, the Toyota powertrain control benchmark \cite{Roohi2017}, and conformance of CPS \cite{Wang2021}. Statistical testing under STL specifications was considered in \cite{salamati2020data,salamati2021data,jackson2021formal,bartocci2013robustness,BARTOCCI2015} where \cite{salamati2020data,salamati2021data,jackson2021formal} combine data-driven and model-based verification techniques, while \cite{bartocci2013robustness,BARTOCCI2015} present a purely data-driven verification technique to estimate probabilities over robustness distributions of the system. 
However, their approaches do not consider risk metrics and do not leverage the robustness values of STL.

%% file: chapter/problem_statement_new_3_30_2022.tex
\section{Problem Statement}
\label{sec:problem_statement}

In this section, we first present two case studies on which we then introduce the risk verification problem. The case studies considered in this paper are an F1/10 car in a hallway and a unmanned underwater vehicle (UUV) tracking a pipeline on the seafloor. After introducing these case studies, we introduce an abstract version of the risk verification problem. 

\subsection{Case Studies}
\label{sec:caseee}

\begin{figure}[ht!]
\centering
\includegraphics[scale=0.5]{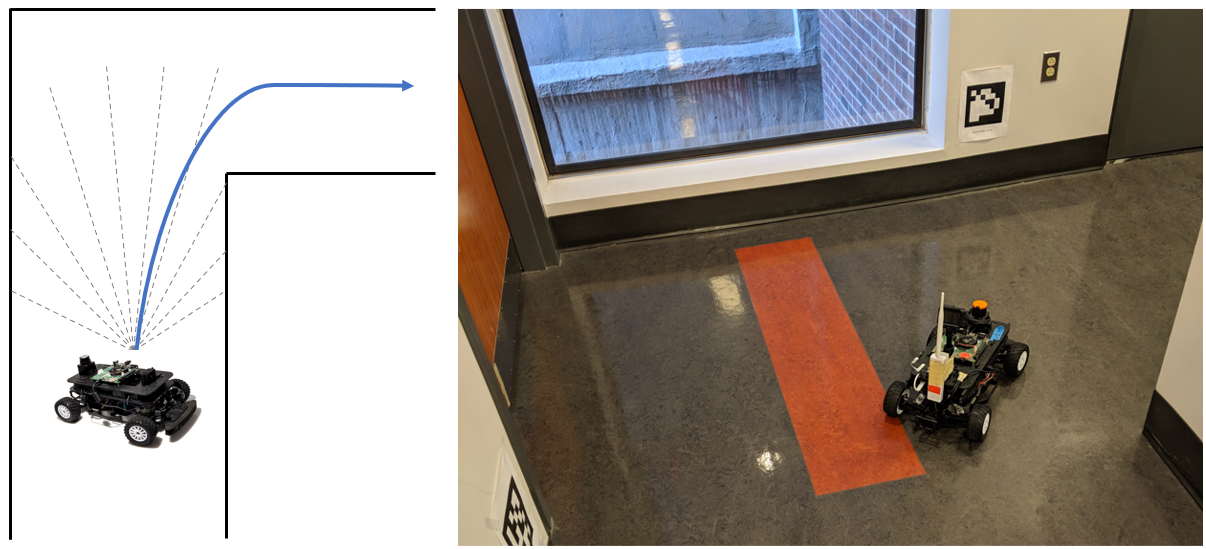}
\caption{The F1/10 operating environment. The grey dashed lines on the left represent LiDAR rays and the blue curve represents a safe path through the hallway.}
\label{fig:f110_fig}
\end{figure}

\textbf{F1/10 car.} The first case study concerns an F1/10 car \cite{F1/10} driving through a hallway with 90-degree turns, as shown in Figure \ref{fig:f110_fig}. The car is equipped with a LiDAR system for sensing the distances to the hallway. We use a standard bicycle model of the car and a ground truth LiDAR model to model the car in simulation. Given this ideal closed-loop simulator setup, we used RL to train NN controllers to safely navigate the car through the hallway. The first question we are interested in is which controller performs best in simulation, i.e., results in the lowest risk. In addition, it has been shown that NN controllers which are verified to be safe in simulation can still show poor performance when evaluated in the real world \cite{ivanov2020}. So if a controller performs well in simulation, what is the risk of it causing the real F1/10 car to crash into the walls of the hallway?

\begin{figure}[ht!]
\centering
\includegraphics[scale=0.25]{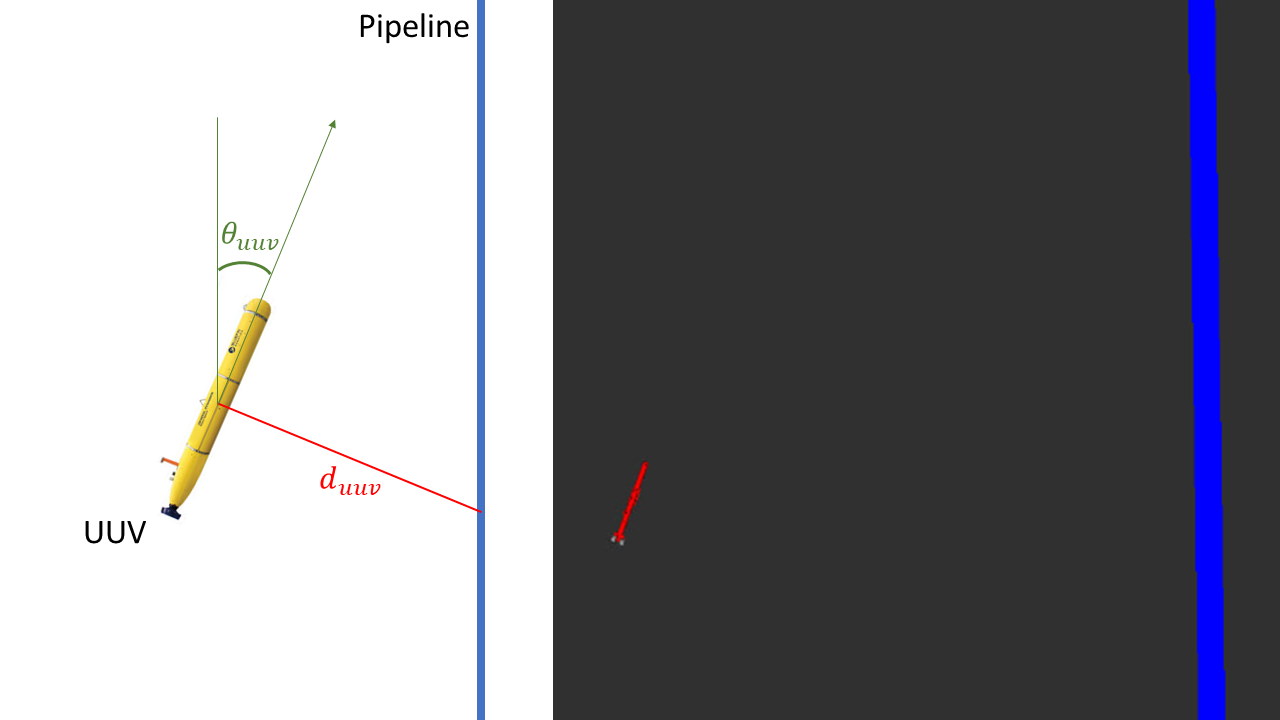}
\caption{The UUV operating environment, both in a schematic drawing and in the high fidelity physics engine. The pipelines are both shown in blue.}
\label{fig:uuv_fig}
\end{figure}

\textbf{UUV.} The second case study concerns a UUV that is tasked with following alongside a pipeline on the seafloor and using a side-scan sonar system to capture and inspect the pipeline, as shown in Figure \ref{fig:uuv_fig}. We consider a high-fidelity UUV model based on the ROS-Gazebo real-time physics engine presented in \cite{Manhaes2016}. As the real-time physics engine is too computationally complex to generate large amounts of data, we fit a partially linearized model of the UUV. Using this model, we use RL to train NN controllers to guide the UUV along the pipe, while keeping an optimal distance for the sonar to accurately inspect the pipe. Once again, we can ask similar questions to the F1/10 case: which NN controller best guides the UUV along the pipe in the modeled environment? What are the risks of these NN controllers not properly guiding the UUV along the pipe in the real-time physics engine?

\subsection{Nominal System Model}
\label{sec:system_model}

The focus in this paper is on the verification of \emph{stochastic dynamic control systems with NN controllers} in the loop. In particular, we are interested in verifying systems in terms of the risk of violating a given specification. As illustrated in the case studies, we distinguish between a (nominal) system model and the actual system (which could be a real system or a high-fidelity simulator with complex dynamics). The system model is useful as it permits theoretical analysis and allows one to use computationally-expensive RL techniques.
%
Thus, we model the nominal system as a standard stochastic dynamic system:
\begin{align}
\begin{split}
    X(t+1)&=f(X(t),u(Y(t)),V(t)), \; X(0)=X_0\\
    Y(t)&=g(X(t),W(t)),
\end{split}
\label{eq:system}
\end{align}
where $X(t) \in \mathbb{R}^n$ is the system state; $Y(t) \in \mathbb{R}^p$ are the observations; the functions $f:\mathbb{R}^n\times\mathbb{R}^m\times\mathbb{R}^v\to\mathbb{R}^n$ and $g:\mathbb{R}^n\times\mathbb{R}^w\to\mathbb{R}^p$ describe the system dynamics and the output measurement map, respectively; the function $u:\mathbb{R}^p\to\mathbb{R}^m$ is an NN controller; and $V(t)\in D_v\subseteq\mathbb{R}^v$ and $W(t)\in D_w\subseteq\mathbb{R}^w$  describe process and measurement noise where the sets $D_v$ and $D_w$ are specified later. 
The initial position $X_0 \in \mathbb{R}^n$ as well as the process and measurement noise $V(t)$ and $W(t)$ are assumed to be random variables so that $X$ defines a stochastic process.

\subsection{Risk Abstraction}
\label{sec:risk_abstraction}

For clarity, in what follows we provide an abstract notion of risk -- the full formalization is presented in Section~\ref{sec:sol_1}. We consider systems subject to real-time safety constraints (e.g., the F1/10 car needs to avoid colliding with walls at all times). One way to express real-time constraints is as follows:
\begin{align}
\label{eq:abstract_risk}
    c(x(t),t)\ge 0 \text{ for all } t\in\mathbb{T}
\end{align}
where $c:\mathbb{R}^n\times \mathbb{T}\to \mathbb{R}$ is a measurable constraint function (e.g., hyper-planes representing walls in the F1/10 example) and $x(t)$ can be a realization of the system trajectory as in~\eqref{eq:system}.\footnote{In Section \ref{sec:STL}, we also consider signal temporal logic (STL) constraints as a more general way of expressing statements close to structured English.} We quantify the robustness of a given trajectory $x(t)$ by measuring how much one would need to alter $x(t)$ to cause it to violate (or satisfy) the constraint $c$. For example, consider a trace of the F1/10 car where the car comes within 1m of colliding with the wall. We would say that this trace has a robustness of 1.

Since the nominal system in~\eqref{eq:system} is stochastic, the robustness value in~\eqref{eq:abstract_risk} is itself a random variable. This is why we are interested in computing risk metrics, e.g., as presented in \cite{Majumdar2020}, over the robustness value to quantify the risk of not satisfying the specification robustly, which we call the \emph{robustness risk}. Commonly used risk metrics are the expected value, the variance, the value-at-risk, or the conditional value-at-risk \cite{rockafellar2000optimization}. We formally introduce risk metrics in Section \ref{sec:sol_1}.

\subsection{Real System}
\label{sec:real_system_model}
Besides estimating the risk of the nominal system in \eqref{eq:system}, we are further interested in quantifying the difference in risk between the nominal system and a system that is different from \eqref{eq:system}, referred to as the real system and denoted as $\overline{X}$. We investigate two ways of describing the relationship between the nominal and the real system, which can then be used to bound the real system's risk with respect to the nominal system. In the first approach, which is standard in the control theory literature, we assume that the real system is captured by the model in~\eqref{eq:system}, but for different distributions of $V$ and $W$, i.e., a perturbed version of~\eqref{eq:system}. However, we also note that this model may not be sufficient to describe a complex real system (or even a high-fidelity simulator). Thus, in the second approach, we analyze how the closeness between the nominal system in \eqref{eq:system} and the real system, potentially very different from \eqref{eq:system}, affects the difference between their risks.

Our main insight is that the difference in risk between these two systems naturally depends on the similarity of the systems and can be bounded in terms of the difference of their system trajectories. Thus, as long as the nominal and real systems are not significantly different, we can provide bounds on the real system's risk with respect to the nominal system. In our experiments, we assume to have access to trajectories from both the nominal and the real system to validate our results.

\subsection{Problem Formulation}
\label{sec:problem_formulation}

We first note that calculating the robustness risk for systems with NN controllers is itself a novel problem. Thus, we address that problem first.
\begin{problem}[Risk Verification]\label{ref:prob1}
Assume the stochastic control system in \eqref{eq:system}
and a measurable constraint function $c:\mathbb{R}^n\times \mathbb{T}\to \mathbb{R}$. Assume a set of NN controllers $\mathcal{U}$ is given. Calculate the robustness risks for the system in \eqref{eq:system} under each NN controller $u_i \in \mathcal{U}$.
\end{problem}
To solve Problem \ref{ref:prob1}, we use data sampled from the system \eqref{eq:system}  under the NN controller $u_i$ to obtain an \emph{upper bound} of the robustness risk with high probability following ideas in our previous work \cite{lindemann2021stl}, but here applied to a closed-loop setting. We emphasize that we are interested in an upper bound of the robustness risk to be risk averse as opposed to being risk neutral by estimating the mean of the robustness risk. Our main goal in Problem \ref{ref:prob1} is to analyze the effect of different risk metrics and different risk levels that can act as a tuning knob between risk averse and risk neutral decision making. We also argue that robustness should be used as a quality measure compared to existing verification techniques that rely on a binary verification answer.

The second problem that we address in this paper is to investigate how the robustness risk, calculated in Problem~\ref{ref:prob1}, can be used to determine how the NN controllers perform on systems that are different from~\eqref{eq:system}, i.e., how the robustness risk can aid in determining if an NN controller is robust enough to be used on the real system. As mentioned above, we consider two versions of this problem: 1) the real system is a perturbed version of the model; 2) the real system dynamics are potentially very different from \eqref{eq:system}.
\begin{problem}[Risk Verification Gap for Perturbed Systems]
\label{ref:prob2New}
Consider the  stochastic control system in~\eqref{eq:system} and a perturbed stochastic process $\overline{X}$ that also follows \eqref{eq:system}, but for different distributions of $V$ and $W$. Two specific cases of \eqref{eq:system} that we consider are Lipschitz continuous and incrementally input-to-state stable systems. We want to find a constant $\Delta\ge 0$ so that the difference in robustness risk between $X$ and $\overline{X}$ is bounded by $\Delta$.
\end{problem}

As already emphasized, we note that the model in~\eqref{eq:system} may not be sufficient to describe a complex real system (or even a high-fidelity simulator) and we may not be able to find $\Delta$ as in Problem \ref{ref:prob2New}. We want to quantify  the risk verification gap of such systems  in terms of the closeness of the systems.
\begin{problem}[Expected Risk Verification Gap for Real System]
\label{ref:prob3}
Consider the  stochastic control system in~\eqref{eq:system} and a real system that may be different from \eqref{eq:system}. Using trajectories drawn from the two systems, find a constant $\Delta \geq 0$ so that the risk verification gap between the two systems is bounded by $\Delta$ with high confidence.
\end{problem}
\begin{remark}
The insights from  Problems~\ref{ref:prob2New} and ~\ref{ref:prob3} also provide a practical way to compare NN controllers on the nominal system. In particular, the controller that achieves the lowest risk on the nominal system is likely to perform best on the real system as well. The experimental evaluation confirms this observation.
\end{remark}

%% file: chapter/risk_background.tex
\section{Risk Verification}
\label{sec:sol_1}

We now present our risk verification framework for solving Problem \ref{ref:prob1}.

\subsection{Random Variables, Risk Measures, and Stochastic Processes}
\label{sec:stoch}

Consider the \emph{probability space} $(\Omega,\mathcal{F},P)$  where $\Omega$ is the sample space, $\mathcal{F}$ is a $\sigma$-algebra of $\Omega$, and $P:\mathcal{F}\to[0,1]$ is a probability measure. More intuitively, an element in  $\Omega$ is an \emph{outcome} of an experiment, while an element in $\mathcal{F}$ is an \emph{event} that consists of one or more outcomes whose probabilities can be measured by the probability measure $P$. 

\begin{figure}[ht!]
\centering
\includegraphics[scale=0.4]{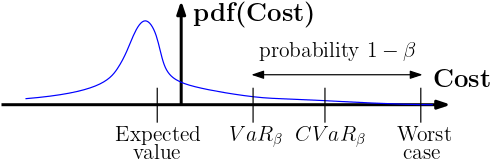}
\caption{Visualization of common risk metrics.}
\label{fig:risk_fig}
\end{figure}

Let $Z$ denote a real-valued \emph{cost random variable}, i.e., a measurable function $Z:\Omega\to\mathbb{R}$. Let $\mathfrak{F}(A,B)$ denote the set of all functions which map input domain $A$ to output domain $B$ (e.g. $Z\in \mathfrak{F}(\Omega,\mathbb{R})$).
A \emph{risk metric} is a function $R:\mathfrak{F}(\Omega,\mathbb{R})\to \mathbb{R}$ that maps from the set of real-valued random variables to the real numbers. The value of $R(Z)$ quantifies the risk associated with cost random variable $Z$.  Risk metrics hence allow for a risk assessment in terms of such cost random variables. Commonly used risk metrics are the expected value, the variance, the value-at-risk, or the conditional value-at-risk \cite{rockafellar2000optimization}, see Figure \ref{fig:risk_fig} for a visualization of these risk metrics.

Particular properties of $R$ that we need in this paper are \emph{monotonicity}, \emph{translational invariance}, and \emph{subadditivity}:
 \begin{itemize}
     \item For two cost random variables $Z,Z'\in \mathfrak{F}(\Omega,\mathbb{R})$, the risk metric $R$ is monotone if $Z(\omega) \leq Z'(\omega)$ for all $\omega\in\Omega$ implies that $R(Z) \le R(Z')$.
     \item For a random variable $Z\in \mathfrak{F}(\Omega,\mathbb{R})$, the risk metric $R$ is translationally invariant if, for any constant $c\in\mathbb{R}$, it holds that $R(Z + c) = R(Z) + c$.
     \item For two cost random variables $Z,Z'\in \mathfrak{F}(\Omega,\mathbb{R})$, the risk metric $R$ is subadditive if  $R(Z + Z') \leq R(Z) + R(Z')$.
 \end{itemize} 

\begin{remark}
In Appendix \ref{app:risk}, we summarize other desirable properties of $R$ such as positive homogeneity, comotone additivity, and law invariance. We also provide a summary of existing risk metrics and emphasize that our presented method is compatible with any of these risk metrics as long as they are monotone, translationally invariant, and subadditive.
\end{remark}

For a time domain $\mathbb{T}\subseteq \mathbb{N}$, a \emph{stochastic process} is a function $X:\mathbb{T}\times \Omega \to \mathbb{R}^n$ where  $X(t,\cdot)$ is a random vector for each fixed $t\in \mathbb{T}$. A stochastic process can   be viewed as a collection of random vectors $\{X(t,\cdot)|t\in \mathbb{T}\}$  that are defined on a common probability space $(\Omega,\mathcal{F},P)$.  For a fixed $\omega\in\Omega$, the function $X(\cdot,\omega)$ is a \emph{realization} of the stochastic process. Note that the stochastic dynamics system in \eqref{eq:system} is a particular form of a stochastic process.

\subsection{The Robustness Risk}
\label{sec:robustness_risk}

Now we formally define constraint functions, constraint safe sets, trace robustness and robustness risk. In this section, we consider specifications as $c(x(t),t)\ge 0 \text{ for all } t\in\mathbb{T}$.

\begin{example} For the F1/10 case study where the car must avoid colliding into the walls of a hallway, we can choose
\begin{align*}
    c(x(t),t):=d_w(x(t)) - \delta
\end{align*} 
where $d_w:\mathbb{R}^n\to \mathbb{R}$ returns the minimum distance from the current car position to the walls and $\delta \in \mathbb{R}_{\geq}$ is the minimum safe distance between the car and the walls. The constraint $c(x(t),t)\ge 0$ is, in this case, equivalent to $d_w(x(t))\ge \delta$ for all $t \in \mathbb{R}_{\geq 0}$.
\end{example}

Importantly, we can quantify how robustly a signal $x$ satisfies the constraint $c(x(t),t)\ge 0$. Therefore, let us define the sets \begin{align*}
    O^c(t)&:=\{x\in\mathbb{R}^n| c(x,t)\ge 0\}\\
    O^{\neg c}(t)&:=\{x\in\mathbb{R}^n| c(x,t)< 0\}
\end{align*}
that indicate the part of the state space where $x(t)$ satisfies and violates the constraint $c(x(t),t)\ge 0$, respectively. We can now define the signed distance $\text{Dist}^c:\mathbb{R}^n\times\mathbb{T}\to\mathbb{R}$ from $x(t)$ to the set $O^c(t)$ as
\begin{align*}
    \text{Dist}^c(x(t),t):=\begin{cases}
    \inf_{x'\in \text{cl}(O^{\neg c}(t))} \|x(t)-x'\| &\text{if } c(x(t),t)\ge 0\\
    -\inf_{x'\in \text{cl}(O^c(t))} \|x(t)-x'\| &\text{otherwise }
    \end{cases}
\end{align*}
where cl$(\cdot)$ denotes the closure operation of a set. In words, $\text{Dist}^c(x(t),t)$ denotes how much one would need to alter the value of signal $x$ at time $t$ to have it violate (satisfy) constraint $c(x(t),t)$, provided $x(t)$ already satisfies (violates) the constraint at time $t$. For instance, a value of $\text{Dist}^c(x(t),t)=2$ means that $x(t)$ would need to be altered by at least $2$ in order to violate $c$ at time $t$ (and vice versa if $\text{Dist}^c(x(t),t)=-2$).

Let us next define the \emph{robustness function} $\rho^c:\mathfrak{F}(\mathbb{T},\mathbb{R}^n)\to\mathbb{R}$, where $\mathfrak{F}(\mathbb{T},\mathbb{R}^n)$ denotes the set of all measurable functions mapping from the domain $\mathbb{T}$ into the domain $\mathbb{R}^n$,   as
\begin{align*}
    \rho^c(x):=\inf_{t\in\mathbb{T}} \, \text{Dist}^c(x(t),t).
\end{align*}
Naturally, notice that $\rho^c(x)\ge 0$ implies that the specification is satisfied by $x$, i.e., that $c(x(t),t)\ge 0$ for all $t\in\mathbb{T}$. If $\rho^c(x)>0$, notice further that all signals $x':\mathbb{T}\to\mathbb{R}^n$ that are such that $\|x(t)-x'(t)\|<\rho^c(x)$ for all $t\in\mathbb{T}$ also satisfy the specification, i.e., are such that $\rho^c(x')>0$. 

In the remainder, we will not evaluate the constraint function $c$ over a deterministic signal $x$, but instead over a stochastic process $X$, i.e., we aim to use $c(X(t,\cdot),t)$ and the robustness of stochastic systems. Therefore, we compute risk metrics of the robustness random variable as $R(-\rho^c(X))$ where we consider $-\rho^c(X)$ as the cost random variable. Note that $-\rho^c(X)$ needs to be a random variable, i.e., $-\rho^c(X(\cdot,\omega))$ needs to be measurable in $\omega$, which we show to hold in Appendix \ref{app:measurability}. 

\subsection{Estimating an Upper Bound of the Robustness Risk}
\label{sec:estimate_risk}
As the system in~\eqref{eq:system} defines a stochastic process, Problem~\ref{ref:prob1} can  be solved by estimating upper bounds of the risk $R(-\rho^c(X))$ using frameworks as presented in \cite{Szrnyi2015,nikolakakis2021quantile,thomas2019concentration}. In this paper, we particularly use the commonly studied risk metrics of the value-at-risk and the conditional value-at-risk. We estimate an upper bound of the robustness risk $R(-\rho^c(X))$ from $N$ observed realizations $X^1,\hdots,X^N$ of the stochastic process $X$. Let us, for convenience, first define the random variable 
\begin{align*}
    Z:=-\rho^c(X).
\end{align*}
For further convenience, let us define the tuple
\begin{align*}
    \mathcal{Z}:=(Z^1,\hdots,Z^N)
\end{align*}
where $Z^i:=-\rho^c(X^i,t)$.

\emph{Value-at-Risk (VaR):}  For a risk level of $\beta\in(0,1)$, the VaR of $Z$ is given by
\begin{align*}
VaR_\beta(Z):= \inf\{\alpha\in\mathbb{R}|F_{Z}(\alpha)\ge \beta\}
\end{align*}
where $F_{Z}(\alpha)$ denotes the CDF of $Z$. To estimate $F_{Z}(\alpha)$, define the empirical CDF 
\begin{align*}
\widehat{F}(\alpha,\mathcal{Z}):=\frac{1}{N}\sum_{i=1}^N \mathbb{I}(Z^i\le \alpha)
\end{align*}
where $\mathbb{I}$ denotes the indicator function. From the empirical CDF, the empirical $VaR_\beta(Z)$ is obtained as follows:
\begin{align*}
    \widehat{VaR}_\beta(\mathcal{Z}) = \text{inf} \left\{ \alpha \in \mathbb{R} | \widehat{F}(\alpha,\mathcal{Z})\ge \beta \right\}
\end{align*}

Now we can compute an upper bound of $VaR_{\beta}(Z)$ as follows:

\begin{proposition}[\cite{Szrnyi2015}]\label{prop:quantile}
    Let $\delta \in (0,1)$ be a probability threshold and $\beta \in (0,1)$ be a risk level. Let $Z^1,\hdots,Z^N$ be $N$ independent copies of the random variable $Z$. Then, with probability at least $1-\delta$ it holds that
    \begin{align*}
        VaR_{\beta}(Z) \leq \widehat{VaR}_{\beta + c_N(\delta)}(\mathcal{Z})
    \end{align*}
    where 
    \begin{align*}
    c_N(\delta) := \sqrt{\frac{1}{2N} \text{log} \Big(\frac{\pi^2 N^2}{3 \delta}}\Big).
\end{align*}
\end{proposition}

\emph{Conditional Value-at-Risk (CVaR):} For a risk level of $\beta\in(0,1)$, the CVaR of $Z$ is given by
\begin{align*}
	    CVaR_\beta(Z):=\inf_{\alpha \in {\mathbb{R}}} \; \alpha+(1-\beta)^{-1}E([Z-\alpha]^+)
\end{align*} 
	where $[Z-\alpha]^+:=\max(Z-\alpha,0)$. For estimating  $CVaR_\beta(Z)$ from data $\mathcal{Z}$, we focus here on the case where the random variable $Z=-\rho^\phi(X,t)$ has bounded support (for a fixed $t$) and leverage results from \cite{thomas2019concentration}. In particular, we assume that $P(Z\le b)=1$ for a known constant $b$. Note that $Z$ has bounded support when $\rho^\phi(X,t)$ is bounded, which can be achieved  either by construction of $\phi$ or by clipping off $\rho^\phi(X,t)$ outside the interval $(-\infty,b]$. We can then compute an upper bound of $CVaR_{\beta}(Z)$ as follows:


\begin{proposition}[\cite{thomas2019concentration}]\label{prop:cvar}
    Let $\delta\in(0,0.5]$ be a probability threshold and $\beta\in(0,1)$ be a risk level. Assume that $P(Z\le b)=1$. Let $Z^1,\hdots,Z^N$ be $N$ independent copies of the random variable $Z$. Then, with probability at least $1-\delta$ it holds that
    \begin{align*}
         CVaR_{\beta}(Z) \leq O_{N+1} - \frac{1}{\beta} \sum_{i=1}^{N} (O_{i+1} - O_{i})  \left[ \frac{i}{N} - \sqrt{\frac{\ln(1/\delta)}{2N}} - (1-\beta) \right]^+  
    \end{align*}
    where $x^+:=\text{max}\{x,0\}$, $O_1,\hdots,O_N$ are the order statistics (i.e. $Z^1,\hdots,Z^N$ sorted in ascending order) and $O_{N+1} := b$.
\end{proposition}

%% file: chapter/risk_gap.tex
\section{The Risk Verification Gap between  Nominal and  Perturbed Systems}
\label{sec:risk_gap}

In this section, we are interested in quantifying
the difference in  risk between two systems and understanding their relation. We want to find an upper bound on the robustness risk of the real system $R(-\rho^c(\overline{X}))$ in terms of the robustness risk of the nominal system $R(-\rho^c(X))$. We would hence like to find a bound on the \emph{risk verification gap} $R(-\rho^c(\overline{X}))-R(-\rho^c(X))$. We will show that the risk verification gap can be bounded by the maximal  error  between the trajectories of the nominal and the perturbed system $\|\overline{X}(t,\omega)- X(t,\omega)\|$. Let us, for now, assume that the trajectory error can be bounded by $\Delta$, i.e., assume that
\begin{align}\label{eq:assumedTraceBound}
    \|\overline{X}(t,\omega)- X(t,\omega)\|\le \Delta
\end{align}
for all realizations $\omega\in\Omega$ and all times $t\in\mathbb{T}$. Later in this section, we present three ways of how one may obtain  $\Delta$. Our first goal is to quantify how the bound $\Delta$ affects the difference in robustness, as measured by the function $\rho^c$, by which system trajectories of \eqref{eq:system} and $\overline{X}$ satisfy the specification. 
\begin{lemma}\label{lemma:111_c}
 Consider the stochastic control systems in \eqref{eq:system} under a controller $u$ and $\overline{X}$, and assume that the trajectory error $\Delta$ in \eqref{eq:assumedTraceBound} is known. Let $c:\mathbb{R}^n\times \mathbb{T}\to \mathbb{R}$ be a measurable constraint function. Then it  holds that
 \begin{align}\label{eq:phi_inequ__c}
  \rho^c(X(\cdot,\omega))-\Delta\le \rho^c(\overline{X}(\cdot,\omega))
\end{align} for each realization $\omega\in\Omega$. 
\begin{proof}
See Appendix \ref{app:lemma111_c}.
\end{proof}
\end{lemma}
Now that we have quantified in Lemma \ref{lemma:111_c}  how close the robustness of the systems in \eqref{eq:system} and $\overline{X}$ will be, we can directly use this result to bound the risk verification gap under the assumption that the risk metric $R$ is monotone and translationally invariant.
\begin{theorem}\label{thm:111_c}
Let the same conditions as in Lemma \ref{lemma:111_c} hold and let the risk metric $R$ be monotone and translationally invariant. Then it holds that 
\begin{align*}
    R(-\rho^c(\overline{X}))\le R(-\rho^c(X))+\Delta.
\end{align*}
\begin{proof} 
See Appendix \ref{app:thm111_c}.
\end{proof}
\end{theorem}
We can further use the risk verification gap to compare controllers in terms of their robustness risk. Specifically, we are interested in determining if a controller that has a lower robustness risk on the nominal system than another controller will also have a lower risk on the perturbed system. From the proofs of Lemma \ref{lemma:111_c} and Theorem \ref{thm:111_c} one can, under the same assumption, immediately see that also the inequality $R(-\rho^c(X))\le R(-\rho^c(\overline{X}))+\Delta$ holds. The next result is a straightforward consequence of these results and hence provided without proof.
\begin{corollary}\label{corrr}
Let the same conditions as in Theorem \ref{thm:111_c} hold. Assume that $R(-\rho^c(X_{u_1}))\le R(-\rho^c(X_{u_2}))-c$ for two controllers $u_1$ and $u_2$. Then it holds that $R(-\rho^c(\overline{X}_{u_1}))\le R(-\rho^c(\overline{X}_{u_2}))$ if $2\Delta\le c$.
\end{corollary}

Recall that the bound on the risk verification gap hinges on knowing the bound $\Delta$. We next present three ways to calculate or estimate $\Delta$.

\textbf{Lipschitz constants. }A first, but potentially conservative approach, is to quantify $\Delta$ in terms of Lipschitz constants of the system. Let us therefore consider that $\overline{X}$ is produced by the perturbed system
\begin{align}
\begin{split}
    \overline{X}(t+1)&=f(\overline{X}(t),u(\overline{Y}(t)),\overline{V}(t)), \; \overline{X}(0)=X_0\\
    \overline{Y}(t)&=g(\overline{X}(t),\overline{W}(t)),
\end{split}
\label{eq:system_real_Lipschitz}
\end{align}
where $f$, $g$, and $u$ are as in \eqref{eq:system}, but where the process and measurement noise is different from $V$ and $W$ and instead described by the random variables $\overline{V}(t)\in D_v \subseteq \mathbb{R}^n$ and $\overline{W}(t)\in D_w \subseteq \mathbb{R}^p$. Additionally, assume that $f$, $g$, and $u$ are Lipschitz continuous with known Lipschitz constants and that the sets $D_v$ and $D_w$ are compact. Note that Lipschitz constants of NN controller $u$ can be calculated as, for instance, described in \cite{fazlyab2019efficient}.

At time $t$, and without any further assumption on the controller $u$, we can upper bound the error $\|\overline{X}(t,\omega)- X(t,\omega)\|$ between the system trajectories of the nominal system \eqref{eq:system} and the perturbed system \eqref{eq:system_real_Lipschitz} for each realization $\omega\in\Omega$ by the function 
\begin{align}\label{eq_Lip}
    \Delta(t+1):=L_{f,1}\Delta(t)+L_{f,2}L_u(L_{g,1}\Delta(t)+L_{g,2}w^*)+L_{f,3}v^*
\end{align}
where $\Delta(0):=0$ and where $L_{f,1}$, $L_{f,2}$, and $L_{f,3}$ are the Lipschitz constants of the first, second, and third arguments of $f$, $L_{u}$ is the Lipschitz constant of $u$, $L_{g,1}$ and $L_{g,2}$ are the Lipschitz constants of $g$, and $v^*$ and $w^*$ are calculated as
\begin{align*}
    v^*:=2\max_{v\in D_v}\|v\| \,\;\text{ and } \,\; w^*:=2\max_{w\in D_w}\|w\|.
\end{align*}
 In other words, it holds that
\begin{align*}
    \|\overline{X}(t,\omega)- X(t,\omega)\|\le \Delta(t).
\end{align*}
The derivation of the bound $\Delta(t)$ is shown in Appendix \ref{app:riskGap}. Note now that the bound $\Delta(t)$ depends on time, and in fact grows with time, so that Theorem \ref{thm:111_c} can not be applied directly. However, when we only consider a bounded time interval over which the constraint function $c$ is evaluated, we can derive a corresponding result. The next result hence follows by a minor modification of Lemma \ref{lemma:111_c} and Theorem \ref{thm:111_c} and is stated without proof.

\begin{corollary}\label{cor:1_c}
Consider the stochastic control systems in  \eqref{eq:system} and \eqref{eq:system_real_Lipschitz} under a controller $u$ and assume that the trajectory error $\Delta(t)$ is as in \eqref{eq_Lip}. Let $c:\mathbb{R}^n\times \mathbb{T}\to \mathbb{R}$ be a measurable constraint function with $T:=\max \{\mathbb{T}\}$ being bounded. Let the risk metric $R$ be monotone and translationally invariant. Then it holds that 
\begin{align*}
    R(-\rho^c(\overline{X}))\le R(-\rho^c(X))+\Delta(T).
\end{align*}
\end{corollary}

As remarked previously, such Lipschitz bounds for $\Delta$ may potentially be conservative. Similar bounds, in a different context, can for instance be obtained by considering Lipschitz continuous risk metrics, see  \cite[Definition 4.2]{huang2021off}.

\textbf{Incrementally input-to-state stable controller. } While still considering the perturbed system in \eqref{eq:system_real_Lipschitz}, we may avoid such conservatism and instead use tools from robust control. Particularly, we can assume that the controller $u$ is such that the system in \eqref{eq:system} is incrementally input-to-state stable (see e.g., \cite{tran2016incremental}). Let $D:=(V,W)$ and $D_D:=D_v\times D_w$, then the system \eqref{eq:system} is incrementally input-to-state stable if there exists a function $\gamma \in \mathcal{K}_{\infty}$\footnote{A function $\gamma : \mathbb{R}_{\geq 0} \rightarrow \mathbb{R}_{\geq 0}$ is said to be of class $\mathcal{K}$ if it is continuous, strictly increasing, and $\gamma(0)=0$. It is of class $\mathcal{K}_{\infty}$  if furthermore  $\text{lim}_{s\rightarrow \infty} \; \gamma(s)=\infty$.} such that 
\begin{align*}
    \|X(t,\omega;D_1)- X(t,\omega;D_2)\|\le  \gamma(\sup_{d_1,d_2\in D} \|d_1-d_2\|)
\end{align*}
for every realization $\omega\in\Omega$ where $X(t,\omega;D_i)$ is the solution to \eqref{eq:system} but where $D$ is replaced with any stochastic process $D_i:\mathbb{N}\times\Omega\to D_D$. While this may seem like a stringent assumption on the system, learning controllers that exhibit iISS or similar stability properties is an active area of research \cite{tu2021closing,Havens2021,boffi2021regret}.

It follows in this case that the error bound between the system trajectories of the nominal system \eqref{eq:system} and the perturbed system in \eqref{eq:system_real_Lipschitz} are independent of $t$, i.e., that
\begin{align*}
    \|\overline{X}(t,\omega)- X(t,\omega)\|\le  \gamma(\sup_{d_1,d_2\in D} \|d_1-d_2\|)
\end{align*}
so that we can set $\Delta:=\gamma(\sup_{d_1,d_2\in D} \|d_1-d_2\|)$ and  simply apply Theorem~\ref{thm:111_c}. 

\begin{remark}
    Note that any stable linear system is iISS. This can be straightforwardly shown by plugging in the system dynamics into the iISS definition.
\end{remark}

\textbf{Stochastic Bounds.} Let us now return back to the generic perturbed system $\overline{X}$ from Section \ref{sec:real_system_model}. Note that we have so far assumed a worst case upper bound $\Delta$ as in \eqref{eq:assumedTraceBound}. Let us now assume that the trajectory error is bounded by a random variable, i.e., there exists $\Gamma:\Omega\to\mathbb{R}$ such that
\begin{align}\label{eq:assumedTraceBound1}
    \|\overline{X}(t,\omega)- X(t,\omega)\|\le \Gamma(\omega)
\end{align}
for each realization $\omega\in\Omega$ and all times $t\in\mathbb{T}$. In this case, it can be easily seen that Lemma \ref{lemma:111_c} still holds while using $\Gamma(\omega)$ instead of $\Delta$. Based on this observation, we can bound the risk verification gap by $R(\Gamma)$ if we assume that the risk metric is subadditive instead of being translationally invariant.
\begin{theorem}\label{thm:222_c}
Consider the stochastic control systems in  \eqref{eq:system} and $\overline{X}$ under a controller $u$ and assume that the trajectory error $\Gamma(\omega)$ in \eqref{eq:assumedTraceBound1} is known. Let $c:\mathbb{R}^n\times \mathbb{T}\to \mathbb{R}$ be a measurable constraint function and let the risk metric $R$ be monotone and subadditive invariant. Then it holds that 
\begin{align*}
    R(-\rho^c(\overline{X}))\le R(-\rho^c(X))+R(\Gamma).
\end{align*}
\begin{proof} 
See Appendix \ref{app:thm222_c}.
\end{proof}
\end{theorem}

We note that the bound in Theorem \ref{thm:222_c} is tighter than the bound in Theorem \ref{thm:111_c}. However, the downside is that one will have to estimate $R(\Gamma)$, which will require having knowledge of trajectories of $\overline{X}$ which one could use to directly calculate $R(-\rho^c(\overline{X}))$. An empirical estimation of $R(\Gamma)$ also requires generating sample pairs from the nominal and real systems that feature the same corresponding realizations $\omega$ across all random variables, which may be possible in a sim2sim but not necessarily possible in a sim2real setting. Nonetheless, this approach offers additional insight and a way of quantifying the risk gap between two systems.

%% file: chapter/STL.tex

\section{Signal Temporal Logic (STL)}
\label{sec:STL}

Signal temporal logic \cite{maler2004monitoring} specifications allow for more expressivity than constraint functions $c(x,t)$ that were discussed so far. We now extend the previously derived results to such specifications. 

\subsection{Risk Verification under Signal Temporal Logic Specifications}
STL specifications are constructed from predicates $\mu:\mathbb{R}^n\to\{\top,\bot\}$ where $\top:=\infty$ and $\bot:=-\infty$ encode logical true and false, respectively. Let us associate an observation map $O^\mu\subseteq \mathbb{R}^n$ with $\mu$ as
\begin{align*}
O^\mu:=\mu^{-1}(\top)
\end{align*}
where $\mu^{-1}(\top)$ denotes the inverse image of $\top$ under $\mu$. The set $O^\mu$ hence denotes the parts of the state space where the predicate $\mu$ is true. Often, predicates are defined via a predicate function $h:\mathbb{R}^n\to\mathbb{R}$ as
\begin{align*}
\mu(\zeta):=\begin{cases}
\top & \text{if } h(\zeta)\ge 0\\
\bot &\text{otherwise}
\end{cases}
\end{align*}
for $\zeta\in\mathbb{R}^n$ so that  $O^\mu=\{\zeta\in\mathbb{R}^n|h(\zeta)\ge 0\}$. The syntax of STL is defined as 
\begin{align}\label{eq:full_STL}
\phi \; ::= \; \top \; | \; \mu \; | \;  \neg \phi \; | \; \phi' \wedge \phi'' \; | \; \phi'  U_I \phi''
\end{align}
where $\phi'$ and $\phi''$ are STL formulas and where $U_I$ is the until operator with $I\subseteq \mathbb{R}_{\ge 0}$. The until operator encodes that $\phi'$ always has to hold until $\phi''$ holds at some point within the interval $I$. The meanings of the negation and conjunction operators $\neg$ and $\wedge$ are `not' and  `and'. One can further define the disjunction, eventually, and always operators as $\phi' \vee \phi'':=\neg(\neg\phi' \wedge \neg\phi'')$,
$F_I\phi:=\top U_I \phi$, and
$G_I\phi:=\neg F_I\neg \phi$, respectively. We provide the semantics of STL \cite{maler2004monitoring} that determine whether or not a signal $x$ satisfies an STL formula $\phi$ at time $t$ in Appendix \ref{app:STL_sem}. An STL formula $\phi$ is bounded if all time intervals $I$ in $\phi$ are bounded. Each bounded STL formula $\phi$ has a formula length $L^\phi$ that defines the minimum length a signal must be in order to  be able to say whether or not it satisfies $\phi$, see Appendix \ref{app:STL_sem} for a formal statement.

\emph{Robustness.} One may be especially interested in the robustness by which a signal $x$ satisfies the STL formula $\phi$ at time $t$.  For this purpose, the authors in \cite{fainekos2009robustness} introduce the robustness degree and in particular the function $\text{dist}^{\neg\phi}:\mathfrak{F}(\mathbb{N},\mathbb{R}^n)\times\mathbb{N}\to\mathbb{R}_{\ge 0}$. Intuitively, $\text{dist}^{\neg\phi}(x,t)$ measures the distance of the signal $x$ to the set of signals that violate $\phi$ at time $t$ (see Appendix \ref{app:STL} for a formal definition). It holds that $\text{dist}^{\neg\phi}(x,t)=0$ indicates that $x$ violates or marginally satisfies $\phi$, while $\text{dist}^{\neg\phi}(x,t)>0$ indicates that $x$ satisfies $\phi$. In fact, larger values of $\text{dist}^{\neg\phi}(x,t)$ indicate that $\phi$ is satisfied more robustly.  Unfortunately, $\text{dist}^{\neg\phi}(x,t)$ is in general difficult to calculate. The  authors in \cite{fainekos2009robustness} introduce the \emph{robust semantics} $\rho^\phi:\mathfrak{F}(\mathbb{N},\mathbb{R}^n)\times \mathbb{N}\to \mathbb{R}$ as a tractable but more conservative robustness estimate, i.e., it holds that $\rho^{\phi}(x,t)\le \text{dist}^{\neg\phi}(x,t)$ \cite[Theorem 28]{fainekos2009robustness}. A formal definition is given in Appendix \ref{app:STL}.  


Instead of computing the STL robustness of deterministic traces, we are interested in evaluating the STL robustness over a stochastic process $X$. We use random variable $\text{dist}^{\phi}(X,t)$ to denote the robustness values of $X$ for specification $\phi$. The risk of the stochastic process $X$ not satisfying $\phi$ robustly, called the ``STL robustness risk'' \cite{lindemann2021stl}, is defined to be $R(-\text{dist}^{\neg\phi}(X,t))$. As previously remarked, the function $\text{dist}^{\neg\phi}$ is in general hard to calculate and work with. In \cite[Thm. 3]{lindemann2021stl}, we show that  \begin{align*}
    R(-\text{dist}^{\neg\phi}(X,t))\le R(-\rho^\phi(X,t))
\end{align*}
if $R$ is a monotone risk metric. In other words, $R(-\rho^\phi(X,t))$ is more risk-averse than $R(-\text{dist}^{\neg\phi}(X,t))$ and may instead be used. Now, by setting $Z:=-\rho^\phi(X,t)$, we can estimate an upper bound to the STL robustness risk in the same way as discussed in Section \ref{sec:estimate_risk}.  


\subsection{The Risk Verification Gap for Signal Temporal Logic}

For STL specifications $\phi$ we can quantify the risk verification gap similarly to Section \ref{sec:risk_gap} where we considered specifications defined by $c(x,t)$. The next Lemma directly resembles Lemma \ref{lemma:111_c}.


\begin{lemma}\label{lemma:1}
Consider the stochastic control systems in  \eqref{eq:system}  under a controller $u$ and $\overline{X}$. Let  $\phi$ be an STL formula in positive normal form\footnote{A formula $\phi$ is in positive normal form if $\phi$ contains no negations, or negations only appear in front of predicates so that they can be absorbed into them. This is without loss of generality as it is known that every STL formula can be re-written in positive normal form \cite{sadraddini2015robust}.} and assume that the trajectory error $\Delta$ in \eqref{eq:assumedTraceBound} is known. Then it  holds that
 \begin{align}\label{eq:phi_inequ}
  \rho^\phi(X(\cdot,\omega),t)-\Delta\le \rho^\phi(\overline{X}(\cdot,\omega),t))
\end{align} 
for each realization $\omega\in\Omega$ and time $t\in\mathbb{N}$. If additionally $\phi$ is a bounded STL formula in positive normal form with formula length $L^\phi$ and the trajectory error $\Delta(t)$ is as in \eqref{eq_Lip}, then it  holds that
 \begin{align}\label{eq:phi_inequ_}
  \rho^\phi(X(\cdot,\omega),t)-\Delta(t+L^\phi)\le \rho^\phi(\overline{X}(\cdot,\omega),t))
\end{align}
for each realization $\omega\in\Omega$ and time $t\in\mathbb{N}$.

\begin{proof}
See Appendix \ref{app:lemma1}.
\end{proof}
\end{lemma}

Based on this lemma, we can state similar results as in Theorem \ref{thm:111_c} and Corollary \ref{cor:1_c} when considering STL specifications $\phi$.

\begin{theorem}\label{thm:1}
Let the conditions of Lemma \ref{lemma:1} hold and let the risk metric $R$ be monotone and translationally invariant. If $\phi$ is an STL formula in positive normal form and the trajectory error $\Delta$ in \eqref{eq:assumedTraceBound} is known, it holds that 
\begin{align*}
    R(-\rho^\phi(\overline{X},t))\le R(-\rho^\phi(X,t))+\Delta.
\end{align*}
If additionally $\phi$ is a bounded STL formula in positive normal form with formula length $L^\phi$ and the trajectory error $\Delta(t)$ is as in \eqref{eq_Lip}, then it holds that
\begin{align*}
    R(-\rho^\phi(\overline{X},t))\le R(-\rho^\phi(X,t))+\Delta(t+L^\phi).
\end{align*}


\begin{proof}
See Appendix \ref{app:thm1}.
\end{proof}
\end{theorem}

%% file: chapter/case_studies.tex
\section{Case Studies}
\label{sec:caseStudies}

\subsection{F1/10}
\label{sec:F1/10CaseStudy}

Recall the F1/10 system described in Section \ref{sec:problem_statement}, where an F1/10 car must travel through a hallway using LiDAR and an NN controller. The car must travel through the hallway without colliding into any walls, which is captured with the following constraint function:
\begin{align}\label{eq:F10spec}
    c(x(t),t):= d(x(t)) > d_{w}
\end{align}

\subsubsection{Nominal System}
\label{sec:F1/10_nomial_system}
We used a standard bicycle model as the model of the car, denoted as $f$. The LiDAR model, denoted as $g$, returns the ground truth value of each of the 21 LiDAR rays (see the grey dashed lines in Figure \ref{fig:f110_fig} for what each LiDAR ray looks like). There is uniform noise $W(t)$ applied to each LiDAR ray's distance. The controller, denoted as $\text{NN}$, takes arrays of LiDAR rays as input and returns steering commands as output. Equation \eqref{eq:f1tenthsystemsimlidar} shows the F1-10th with simulated LiDAR written in the form of \eqref{eq:system}. The state is $X = \begin{bmatrix} x & y & v & \theta \end{bmatrix}^T$, where $x$ and $y$ are the car's position, $v$ is the car's velocity, and $\theta$ is the car's heading. For the full model description, see \cite{ivanov2020}.
\begin{subequations}\label{eq:f1tenthsystemsimlidar}
\begin{align}
    X(t+1)&=f(X(t),\text{NN}(Y(t))), \; X(0)=X_0\\
    Y(t)&=g(X(t),W(t)),
\end{align}
\end{subequations}

\subsubsection{Perturbed System}
\label{sec:F1/10_perturbed_system}
A common problem with LiDAR is that the rays can bounce off reflective surfaces, causing the sensor to give spuriously large distance readings. To address this issue, we created two more LiDAR models, one with randomly dropped LiDAR rays (denoted by $g_d$) and one where we learn the observation model from data using a generative adversarial network (GAN) (denoted by $g_{gan}$). For the dropped ray model, at the start of each new trace we randomly select  $5$ LiDAR rays to get dropped (thus modeling random reflections) for the entire trace, as opposed to the noisy ground truth distance. This is a crude model of how rays are reflected, since in the real world each LiDAR ray can be reflected and recovered over the course of a trace and the environmental conditions affect when rays get dropped. 

Alternatively, the GAN model aims to capture LiDAR artifacts that are observed in a specific environment. To train the GAN model, we used real data collected in our prior work~\cite{ivanov2020}. Specifically, the GAN takes the car's (estimated) state as input and outputs a distribution of the LiDAR rays observed from that position. We used the Wasserstein GAN training algorithm~\cite{arjovsky2017,gulrajani2017} with a deconvoluational NN architecture \cite{radford2016}. In Section \ref{sec:ganAnalysis} we perform a quantitative analysis of the GAN.


\subsubsection{Controller Training}
\label{sec:F1/10_controller_training}

Next, we used the system in \eqref{eq:f1tenthsystemsimlidar} to train 6 different NN controllers for navigating the car through the hallway, taking LiDAR rays as input and giving steering commands as output. To explore the effect of different RL algorithms and different architectures, three of the NNs were trained using the Deep Deterministic Policy Gradient (DDPG) algorithm \cite{Lillicrap2016ContinuousC} and the other three were trained using the Twin Delayed DDPG (TD3) algorithm~\cite{Fujimoto2018}. The reward function is borrowed from \cite{ivanov2020}; it includes a large penalty for the car hitting the wall to enforce learning safe controllers. Each NN consists of two layers with tanh activations with either 64 neurons each (denoted by $64 \times 64$) or 128 neurons each (denoted by $128 \times 128$). In summation, we used DDPG to train two $64 \times 64$ NNs and one $128 \times 128$ NN and we used TD3 to train two $64 \times 64$ NNs and one $128 \times 128$ NN. Note that using the same architecture and training algorithm leads to different NNs due to randomness in the training procedure.

\subsubsection{GAN Evaluation}
\label{sec:ganAnalysis}

To determine how realistic the simulated F1/10 car with the GAN LiDAR model is, we are interested in experimentally determining the quality of the GAN that we trained to emulate real LiDAR readings.
A common method for assessing the quality of GANs is through inspecting the behavior of whatever component takes the GAN output as input. For example, the inception score (IS) metric grades image generating GANs by running the GAN images through an image classifier and examining the difference between the classifier's accuracy on real versus generated images \cite{salimans2016}. These indirect methods are popular because it is difficult to directly quantify what a good (high-dimensional) GAN output should look like. Thus, we compare the control commands issued by the controllers under the three different LiDAR models with the control commands issued when using real LiDAR data as inputs to the controllers.

An additional challenge in our case is that we would like the GAN to produce realistic data for each system state. Since we do not have ground truth states for the real F1/10 data, we compare the overall distributions of the controller commands in the four different settings (simulated, missing ray, GAN, and real Lidar data). Specifically, we use the initial straight portion of the hallway, since that part of the hallway should give similar LiDAR readings and requires similar control commands to navigate. Although we do not have ground truth positions for the real F1/10 runs, the first half of each (seven-second) run roughly corresponds to the straight segment before the turn. Thus, we conservatively use only the first three seconds of LiDAR readings for each run of the real F1/10 car, in order to compute the control input distributions in the straight segment.

\begin{figure}[t!] 
\begin{subfigure}{0.32\textwidth}
\includegraphics[width=\linewidth]{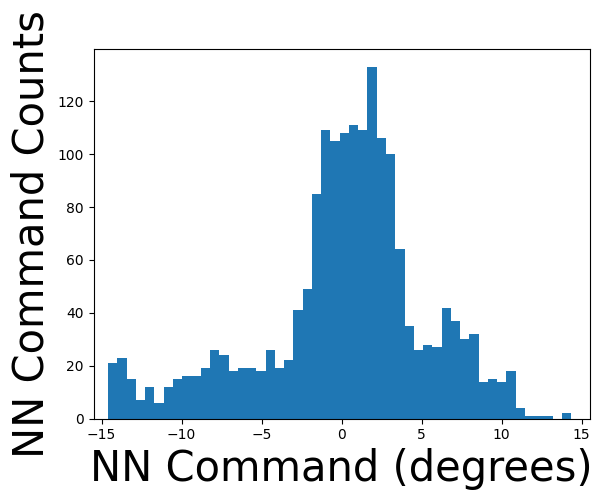}
\caption{Controller 4 Missing Ray} \label{fig:f1-10MissingRayCommandsC4}
\end{subfigure}\hspace*{\fill}
\begin{subfigure}{0.32\textwidth}
\includegraphics[width=\linewidth]{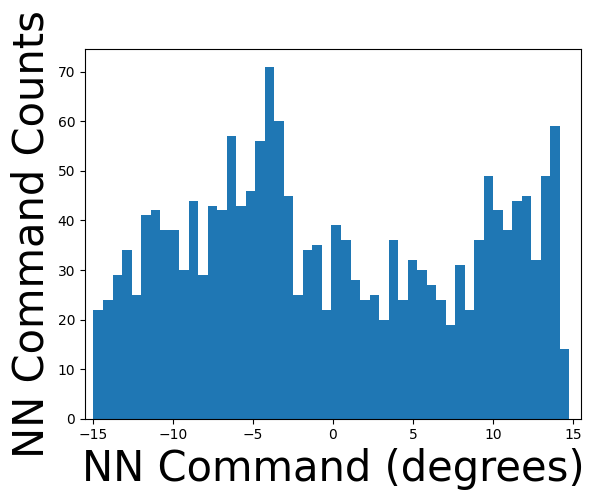}
\caption{Controller 4 GAN} \label{fig:f1-10ganCommandsC4}
\end{subfigure}\hspace*{\fill}
\begin{subfigure}{0.32\textwidth}
\includegraphics[width=\linewidth]{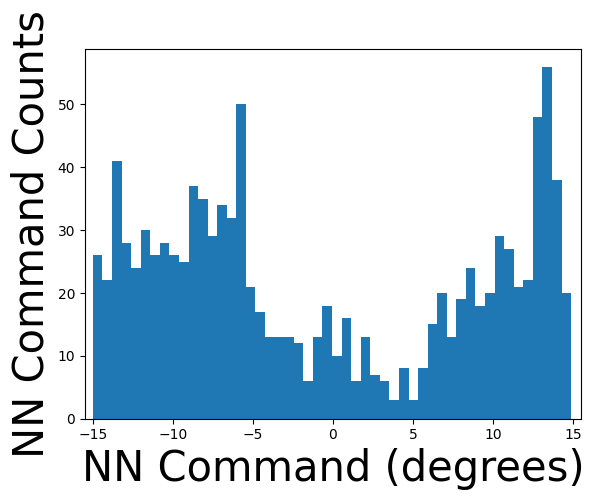}
\caption{Controller 4 Real LiDAR} \label{fig:f1-10realLiDARCommandsC4}
\end{subfigure}
\medskip
\begin{subfigure}{0.32\textwidth}
\includegraphics[width=\linewidth]{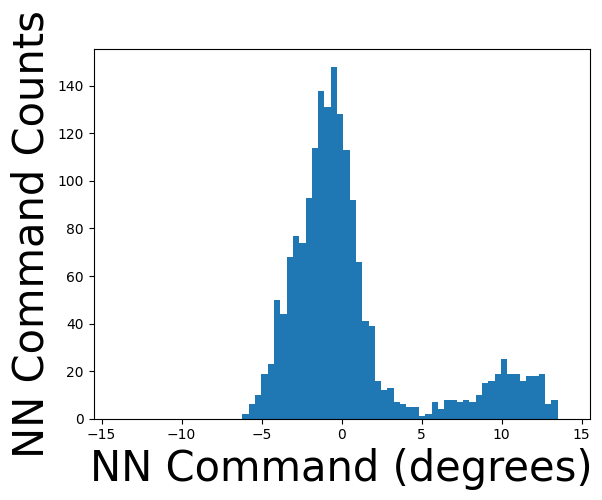}
\caption{Controller 5 Missing Ray} \label{fig:f1-10MissingRayCommandsC5}
\end{subfigure}\hspace*{\fill}
\begin{subfigure}{0.32\textwidth}
\includegraphics[width=\linewidth]{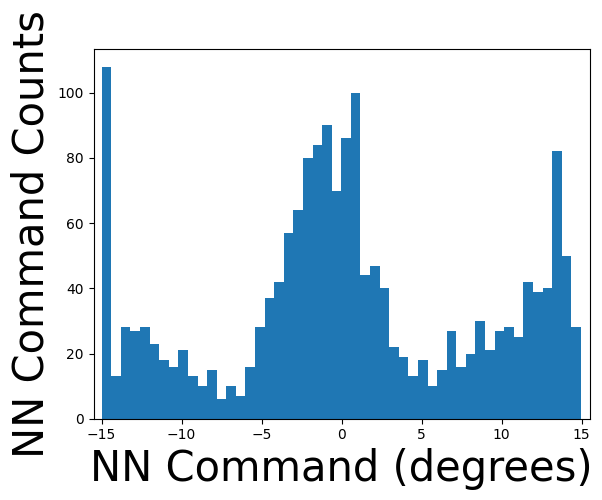}
\caption{Controller 5 GAN} \label{fig:f1-10ganCommandsC5}
\end{subfigure}\hspace*{\fill}
\begin{subfigure}{0.32\textwidth}
\includegraphics[width=\linewidth]{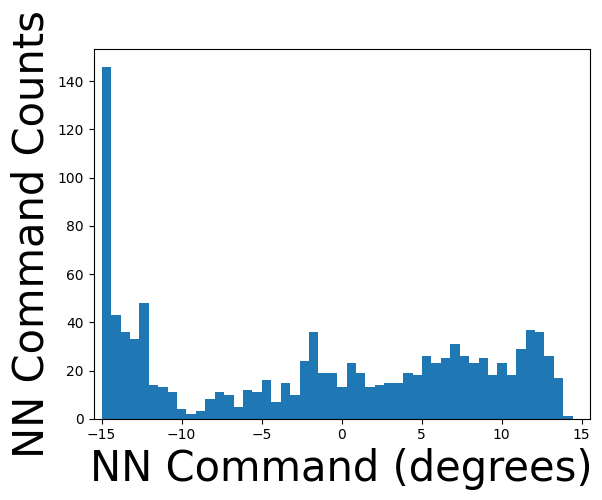}
\caption{Controller 5 Real LiDAR} \label{fig:f1-10realLiDARCommandsC5}
\end{subfigure}
\medskip

\caption{Histograms showing control command distributions for the dropped ray LiDAR model, GAN LiDAR model, and real LiDAR data for various controllers.} \label{fig:f1-10GanRealLiDARHists_1}
\end{figure}

First, we compute the Wasserstein distance between the distributions of control commands for the three types of LiDAR models with the distribution of the control commands on the real LiDAR data, which is shown in Table \ref{Tab:GanDistTable}. The Wasserstein distance is a standard metric for the distance between distributions -- intuitively, it measures the probability mass that needs to be moved in order to convert one distribution into a another \cite{vallender1974}. Across all six controllers, the GAN LiDAR model controls have a considerably smaller Wasserstein distance to the real LiDAR controls than either of the other two LiDAR models do. This is evidence that the GAN does indeed do a reasonable job of approximating the real LiDAR data. Next, we plot histograms of the control commands for the GAN LiDAR and real LiDARs for all six controllers. The histograms of controllers 4 and 5 are in Figure \ref{fig:f1-10GanRealLiDARHists_1} and the histograms of the other 4 controllers are in \ref{app:GANFigures}. The histograms show similar patterns of control commands for most of the controllers. For example, controller 4 shows high levels of actions on $[10,15]$ and $[-10,-15]$ and few actions from $[-3,5]$ for both the GAN and real LiDAR. However, the GAN and real LiDAR histograms for controller 5 are quite different from one another, which demonstrates that the GAN is not a perfect approximation of the real LiDAR.

These results demonstrate that the GAN results in a closer approximation of the real LiDAR than either of the other two LiDAR models we considered. This opens up future research directions of using GANs to make simulations more realistic.

\begin{table*}[h]
\caption{Wasserstein distances of control commands between different LiDAR models and real LiDAR data. The best (lowest) value in each row has been bolded.}
    \centering
    \begin{tabular}{|c|c|c|c|c|}
    \hline \specialcell{Controller \\ Name} & \specialcell{Controller \\ Type} & \specialcell{Simulated LiDAR\\ to Real LiDAR}  & \specialcell{Missing Ray LiDAR\\ to Real LiDAR}    & \specialcell{GAN LiDAR \\ to Real LiDAR} \\ \hline
    1 & DDPG $64 \times 64$ & 3.5173 & 3.4613 & \textbf{1.0903} \\ \hline
    2 & TD3 $64 \times 64$ & 7.1850 & 7.4398 & \textbf{2.2854} \\ \hline
    3 & DDPG $128 \times 128$ & 2.0057 & 1.9783 & \textbf{0.9140} \\ \hline
    4 & DDPG $64 \times 64$ & 5.6847 & 5.4686 & \textbf{1.7944} \\ \hline
    5 & TD3 $128 \times 128$ & 5.7271 & 6.1513 & \textbf{2.3520} \\ \hline
    6 & TD3 $64 \times 64$ & 7.5859 & 7.8669 & \textbf{1.2940} \\
    \hline
    \end{tabular}
    \label{Tab:GanDistTable}
\end{table*}

\subsubsection{Experiments}
\label{sec:F1/10_experiments}

We are now interested in computing the robustness risk of the various NN controlled F1/10 systems satisfying the specification in \eqref{eq:F10spec}. To do so, we first generate 50000 traces for each NN and for each of the three LiDAR models. We simulate the right-hand turn scenario used in prior work~\cite{ivanov2020}. From these data, we compute the VaR and CVaR risks for each of the controllers using a single risk level of $\beta:=0.9$ and confidence of $1-\delta:=0.95$, which are shown in Tables \ref{Tab:f1tenthVars} and \ref{Tab:f1tenthCVars}. The least risky controller is indicated in bold font.

\begin{table*}[h]
\caption{F1/10 VaR values for different LiDAR models under the specification given in \eqref{eq:F10spec}.}
    \centering
    \begin{tabular}{|c|c|c|c|c|}
    \hline \specialcell{Controller \\ Name}  & \specialcell{Controller \\ Type}  & \specialcell{Simulated \\ LiDAR}  & \specialcell{Dropped Ray\\ LiDAR}    & \specialcell{GAN \\ LiDAR} \\ \hline
    1 & DDPG $64 \times 64$ & \textbf{-0.0907} & 0.1113 & 0.1222 \\ \hline
    2 & TD3 $64 \times 64$ & -0.0589 & 0.0840 & 0.0853 \\ \hline
    3 & DDPG $128 \times 128$ & -0.0366 & 0.1649 & 0.1188 \\ \hline
    4 & DDPG $64 \times 64$ & -0.0905 & \textbf{0.0263} & 0.1060 \\ \hline
    5 & TD3 $128 \times 128$ & -0.0668 &0.0413 & 0.0801 \\ \hline
    6 & TD3 $64 \times 64$ & -0.0498 & 0.0300 & \textbf{0.0545} \\
    \hline
    \end{tabular}
    \label{Tab:f1tenthVars}
\end{table*}

\begin{table*}[h]
\caption{F1/10 CVaR values for different LiDAR models under the specification given in \eqref{eq:F10spec}.}
    \centering
    \begin{tabular}{|c|c|c|c|c|}
    \hline \specialcell{Controller \\ Name}  & \specialcell{Controller \\ Type}  & \specialcell{Simulated \\ LiDAR}  & \specialcell{Dropped Ray\\ LiDAR}    & \specialcell{GAN \\ LiDAR} \\ \hline
    1 & DDPG $64 \times 64$ & \textbf{-0.0802} & 0.1518 & 0.1666 \\ \hline
    2 & TD3 $64 \times 64$ & -0.0436 & 0.1259 & 0.1213 \\ \hline
    3 & DDPG $128 \times 128$ & -0.0303 & 0.2008 & 0.1840 \\ \hline
    4 & DDPG $64 \times 64$ & -0.0791 & 0.1064 & 0.1521 \\ \hline
    5 & TD3 $128 \times 128$ & -0.0562 & 0.0945 & 0.1059 \\ \hline
    6 & TD3 $64 \times 64$ & -0.0435 & \textbf{0.0782} & \textbf{0.1032} \\
    \hline
    \end{tabular}
    \label{Tab:f1tenthCVars}
\end{table*}

First, note that all controllers perform best on the simulated LiDAR model. In fact, the risk values are all negative on the simulated LiDAR model. This indicates that the controllers are all generally safe for the simulated LiDAR model, as negative risk robustness values indicate that the system is safe, while positive ones indicate that the system is unsafe. This makes sense as that is the model on which the controllers were trained and is consistent with the risk verification gap theory. We also see that the controllers perform better on the dropped ray LiDAR model than the GAN LiDAR model. That is because the dropped ray LiDAR model is more similar to the simulated LiDAR model than the GAN model is. Note that this is consistent with the risk verification gap theory presented in Section \ref{sec:risk_gap}. 

Note next that there is little correlation between which controllers perform well on the simulated LiDAR model and which perform well on the dropped ray and GAN LiDAR models. This highlights the potential conservatism of the risk verification gap theory as per Corollary \ref{corrr}, which is that if the nominal model is quite different from the perturbed model, then the bound on the risk gap $\Delta$ may be large (this is analyzed in more detail in the next section). We remark that this discrepancy is smaller for the second case study. However, controllers that perform well on the dropped ray model appear to also perform well on the GAN LiDAR model. 

\begin{figure}
\centering
\includegraphics[scale=0.5]{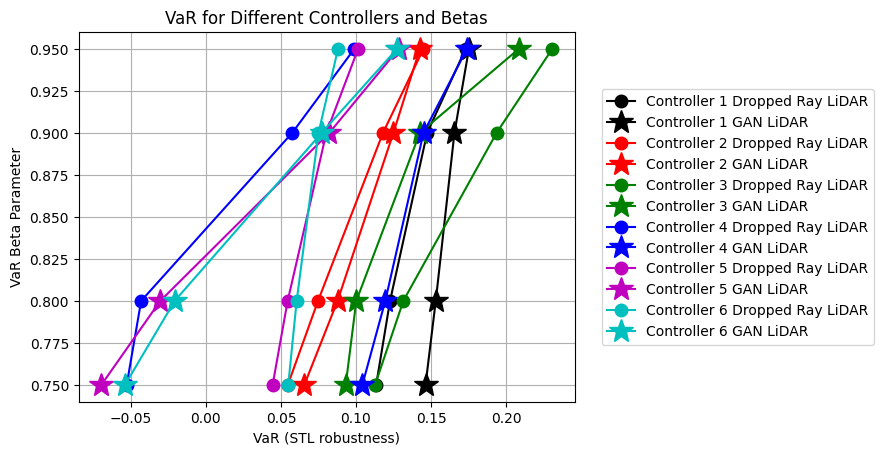}
\caption{Value at Risks (VaRs) for the F1/10 car under the specification given in \ref{eq:F10spec} for the dropped ray and GAN LiDAR models and various controllers.}
\label{fig:f1-10VarManyBetas}
\end{figure}

\begin{figure}
\centering
\includegraphics[scale=0.5]{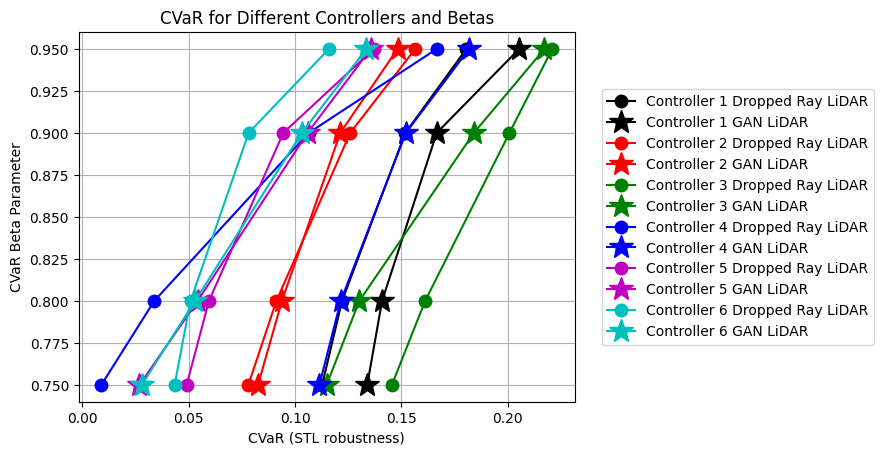}
\caption{Conditional Value at Risks (CVaRs) for the F1/10 car under the specification given in \ref{eq:F10spec} for the dropped ray and GAN LiDAR models and various controllers.}
\label{fig:f1-10CVarManyBetas}
\end{figure}

Next, we varied the $\beta$ values for the VaR and CVaR metrics. These $\beta$ values control how much or little of the tail gets considered in the risk. Larger values only consider the extremes of the tails. Figures \ref{fig:f1-10VarManyBetas} and \ref{fig:f1-10CVarManyBetas} show the different risk values for various controllers for both the dropped ray and GAN LiDAR models. 
Note that we still observe that the controllers which perform better on the dropped ray LiDAR also perform well on the GAN LiDAR.
In particular, for the dropped ray LiDAR Controller 4 shows very good CVaR values for smaller $\beta$ values, but as $\beta$ increases the performances becomes dominated by other controllers. Controller 4 then performs much worse than those controllers on the GAN data. This illustrates the benefit of using risk metrics analyzing systems in simulation. The poor performance of Controller 4 at the extreme tails of its robustness distribution properly indicates that it would not generalize well to the GAN LiDAR model.

\begin{figure}[t!] 
\begin{subfigure}{0.32\textwidth}
\includegraphics[width=\linewidth]{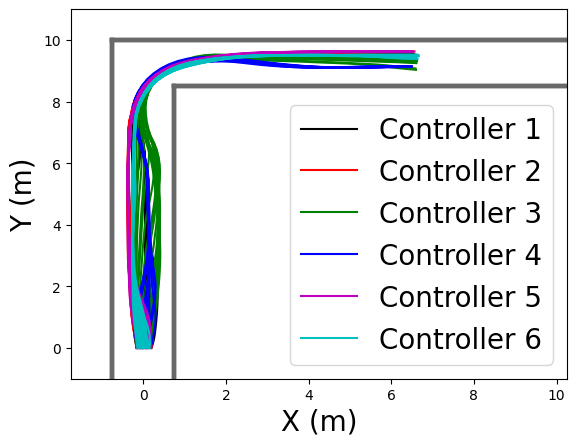}
\caption{Simulated LiDAR traces} \label{fig:f1-10TracesSimLidar}
\end{subfigure}
\begin{subfigure}{0.32\textwidth}
\includegraphics[width=\linewidth]{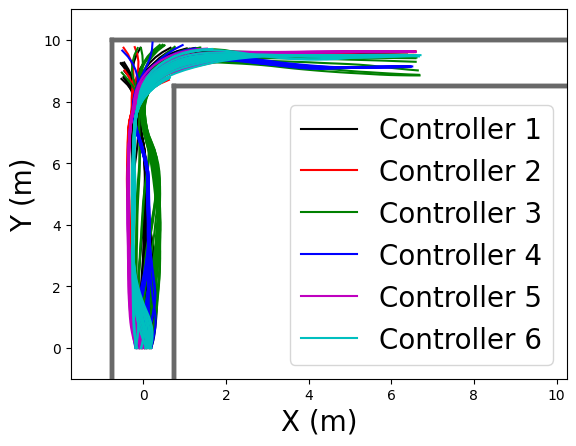}
\caption{Missing ray LiDAR traces} \label{fig:f1-10TracesMissingRay}
\end{subfigure}
\begin{subfigure}{0.32\textwidth}
\includegraphics[width=\textwidth]{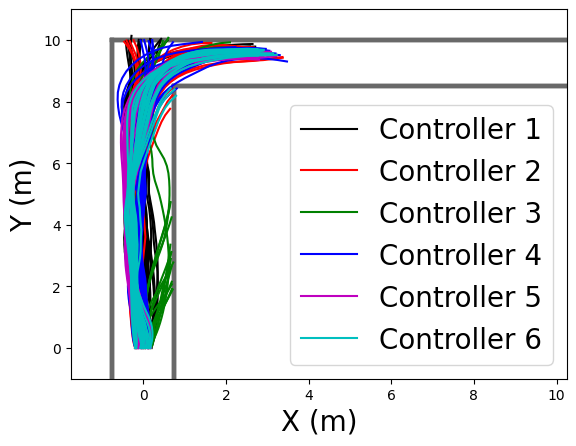}
\caption{GAN LiDAR traces} \label{fig:f1-10TracesGAN}
\end{subfigure}

\caption{Traces for different LiDAR models and controllers.} \label{fig:f1-10Traces}
\end{figure}

Finally, we plot $20$ traces at random for each controller for both the missing ray and GAN LiDAR models. These traces can be seen in Figure \ref{fig:f1-10Traces}. As expected, controller 6 crashes in the walls the least. However, note that the controllers are in general unsafe, especially for the GAN LiDAR model. This is reflected in the robustness risks being positive in Tables \ref{Tab:f1-10TraceDiffVaRTable} and \ref{Tab:f1-10TraceDiffCVaRTable}, i.e., using robustness risk can indicate the quality of safety.

\subsubsection{Trace Difference Risk Bounds}
\label{sec:F1/10_traceDiffBounds}

In this section, we use the results from Theorems \ref{thm:111_c} and \ref{thm:222_c} to bound the robustness risks of the GAN LiDAR model based on the robustness risks of the dropped ray LiDAR models and the trace differences between the two models. We can define the trace difference bound $\Delta(\omega)$ as
\begin{align}\label{eq:traceDiffF1/10}
     \Delta(\omega) := \sup_{t \in \mathbb{T}} \|\overline{X}(t,\omega)- X(t,\omega)\|
\end{align}
where $X$ denotes the dropped ray LiDAR model and $\overline{X}$ denotes the GAN LiDAR model. The goal of these experiments is to experimentally determine bounds on \eqref{eq:traceDiffF1/10} to use and verify Theorems \ref{thm:111_c} and \ref{thm:222_c}.

\begin{figure}[t!] 
\begin{subfigure}{0.32\textwidth}
\includegraphics[width=\linewidth]{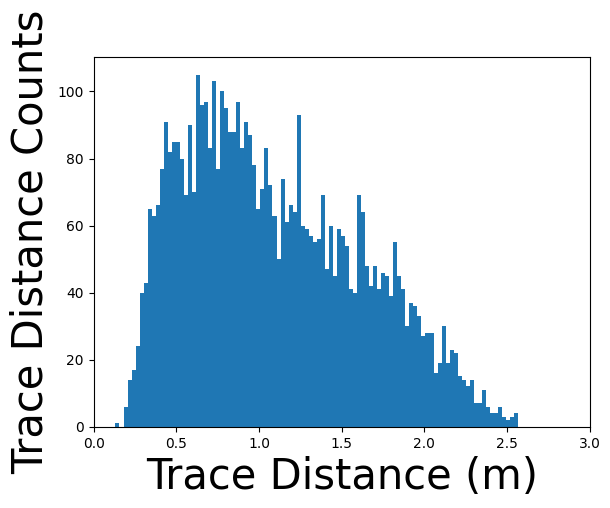}
\caption{Controller 1} \label{fig:f1-10TraceDiffsc1}
\end{subfigure}
\begin{subfigure}{0.32\textwidth}
\includegraphics[width=\linewidth]{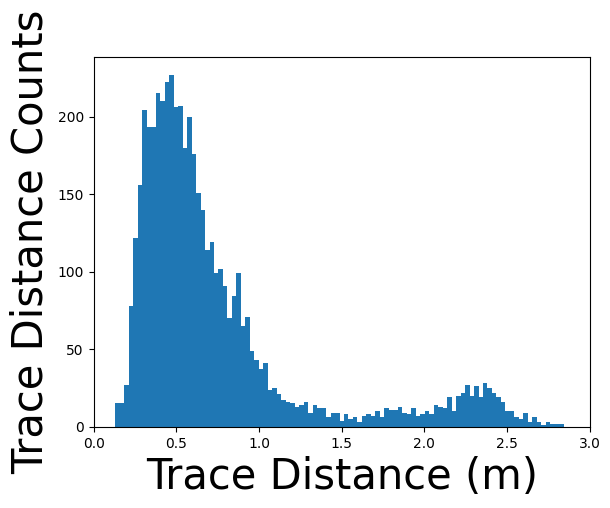}
\caption{Controller 2} \label{fig:f1-10TraceDiffsc2}
\end{subfigure}
\medskip
\begin{subfigure}{0.32\textwidth}
\includegraphics[width=\linewidth]{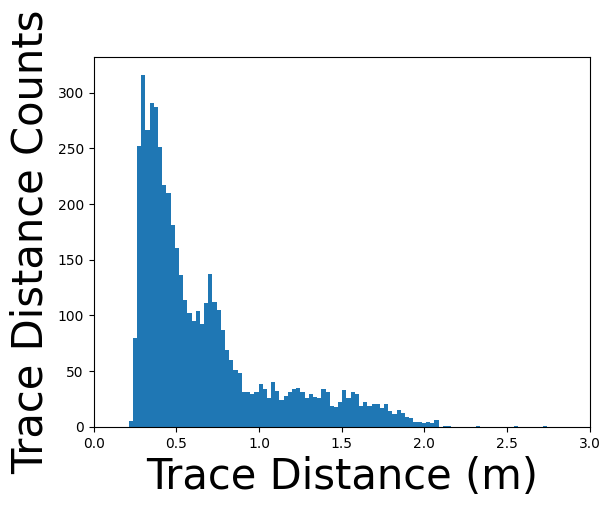}
\caption{Controller 3} \label{fig:f1-10TraceDiffsc3}
\end{subfigure}
\begin{subfigure}{0.32\textwidth}
\includegraphics[width=\linewidth]{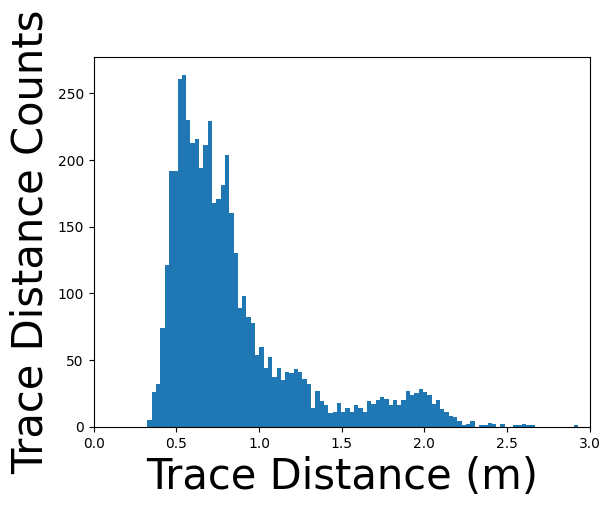}
\caption{Controller 4} \label{fig:f1-10TraceDiffsc4}
\end{subfigure}
\begin{subfigure}{0.32\textwidth}
\includegraphics[width=\linewidth]{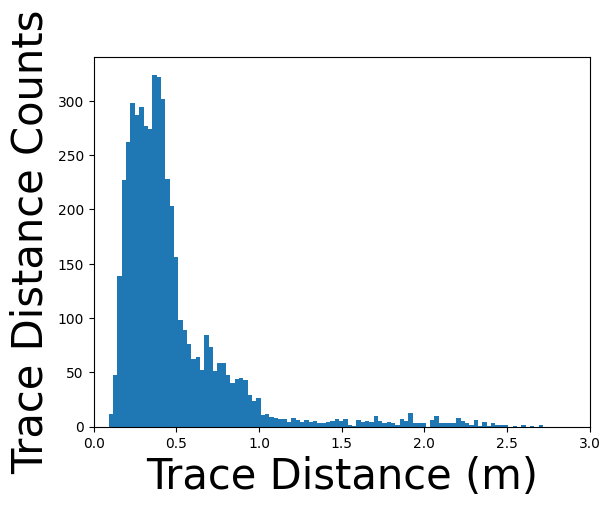}
\caption{Controller 5} \label{fig:f1-10TraceDiffsc5}
\end{subfigure}
\begin{subfigure}{0.32\textwidth}
\includegraphics[width=\linewidth]{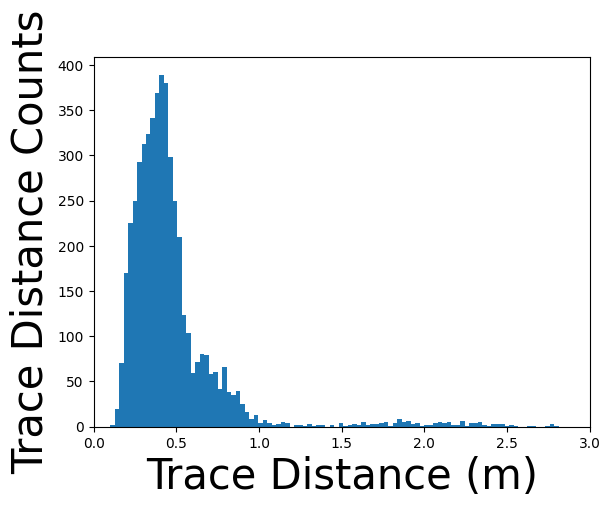}
\caption{Controller 6} \label{fig:f1-10TraceDiffsc6}
\end{subfigure}

\caption{Trace Difference histograms between the dropped ray and GAN LiDAR models for various controllers.} \label{fig:f1-10TraceDiffs}
\end{figure}

To do this, we run both the dropped LiDAR and GAN LiDAR models from the same random initial conditions for $5000$ trials to get a distribution of trace differences, which are shown as histograms in Figure \ref{fig:f1-10TraceDiffs}. Then we compute the supremum, VaR, and CVaR of these trace differences, which we denote as $\sup_{\omega \in \Omega}\Delta(\omega)$, $VaR_\beta({\Delta(\omega)})$, and $CVaR_\beta(\Delta(\omega))$. We refer to these values as the trace difference risks. Finally, we use Theorems \ref{thm:111_c} and \ref{thm:222_c} to combine these trace difference risks with the dropped ray LiDAR model robustness risks to determine upper bounds on the GAN LiDAR model robustness risks. These bounds, along with the trace difference risks, are shown in Tables \ref{Tab:f1-10TraceDiffVaRTable} and \ref{Tab:f1-10TraceDiffCVaRTable}. The GAN risk bounds are all larger than the empirical GAN robustness risks we computed in Tables \ref{Tab:f1tenthVars} and \ref{Tab:f1tenthCVars}, which shows that the bounds are sound. In addition, the relative values of the Theorem \ref{thm:222_c} bounds match those of the empirical GAN robustness risks from Tables \ref{Tab:f1tenthVars} and \ref{Tab:f1tenthCVars}. In fact, the two controllers with the best estimated GAN robustness risks, controllers $5$ and $6$ (see Tables \ref{Tab:f1tenthVars} and \ref{Tab:f1tenthCVars}), also have the best robustness risks bounds when using Theorem \ref{thm:222_c}. This indicates that the bound in Theorem \ref{thm:222_c} is useful in practice. One drawback is that the bounds are a bit larger than the empirical robustness risks, which indicates that the bounds can be conservative. On the other hand, the bounds from using Theorem \ref{thm:111_c} have less of a correlation with the relative robustness risks of the controllers. This highlights the benefit of treating the trace differences as random variables, as opposed to simply considering their upper bounds.

\begin{table*}[h]
\caption{GAN LiDAR VaR bounds estimated from trace bounds and the dropped ray LiDAR VaR bounds.}
    \centering
    \begin{tabular}{|c|c|c|c|c|c|}
    \hline \specialcell{Controller \\ Name} & \specialcell{Dropped Ray\\ LiDAR VaR} & \specialcell{$\sup_{\omega \in \Omega}(\Delta(\omega))$} & \specialcell{$VaR(\Delta(\omega))$} & \specialcell{GAN LiDAR \\ VaR Bound \\ Theorem \ref{thm:111_c}} & \specialcell{GAN LiDAR \\ VaR Bound \\ Theorem \ref{thm:222_c}} \\ \hline
    1 & 0.1113 & \textbf{2.5671} & 1.9074 & \textbf{2.6784} & 2.0187 \\ \hline
    2 & 0.0840 & 2.8430 & 1.7495 & 2.927 & 1.8335 \\ \hline
    3 & 0.1649 & 2.7444 & 1.4020 & 2.9093 & 1.5669 \\ \hline
    4 & \textbf{0.0263} & 2.9291 & 1.5601 & 2.9554 & 1.5864 \\ \hline
    5 & 0.0413 & 2.7183 & 0.8942 & 2.7596 & 0.9355 \\ \hline
    6 & 0.0300 & 2.8128 & \textbf{0.7815} & 2.8428 & \textbf{0.8115} \\
    \hline
    \end{tabular}
    \label{Tab:f1-10TraceDiffVaRTable}
\end{table*}

\begin{table*}[h]
\caption{GAN LiDAR CVaR bounds estimated from trace bounds and the dropped ray LiDAR CVaR bounds.}
    \centering
    \begin{tabular}{|c|c|c|c|c|c|}
    \hline \specialcell{Controller \\ Name}  & \specialcell{Dropped Ray\\ LiDAR CVaR}  & \specialcell{$\sup_{\omega \in \Omega}(\Delta(\omega))$} & \specialcell{$CVaR(\Delta(\omega))$}    & \specialcell{GAN LiDAR \\ CVaR Bound \\ Theorem \ref{thm:111_c}} & \specialcell{GAN LiDAR \\ CVaR Bound \\ Theorem \ref{thm:222_c}} \\ \hline
    1 & 0.1518 & \textbf{2.5671} & 2.2008 & \textbf{2.7189} & 2.3526 \\ \hline
    2 & 0.1259 & 2.8430 & 2.4045 & 2.9689 & 2.5304 \\ \hline
    3 & 0.2008 & 2.7444 & 1.8608 & 2.9452 & 2.0616 \\ \hline
    4 & 0.1064 & 2.9291 & 2.1220 & 3.0355 & 2.2284 \\ \hline
    5 & 0.0945 & 2.7183 & 1.6597 & 2.8128 & 1.7542 \\ \hline
    6 & \textbf{0.0782} & 2.8128 & \textbf{1.5475} & 2.8910 & \textbf{1.6257} \\
    \hline
    \end{tabular}
    \label{Tab:f1-10TraceDiffCVaRTable}
\end{table*}

\subsection{UUV}
\label{sec:UUVCaseStudy}

Recall the UUV system described in Section \ref{sec:problem_statement}, where a UUV is tasked with tracking a pipeline on the seafloor using a side scan sonar and an NN controller. As the sonar works best in a certain range of distances from the pipeline, we consider the following requirement in STL that expresses expeditiousness:
\begin{align} \label{eq:UUVspec}
    \Phi_{UUV}:= &G \left[d_{uuv}<d_l \implies F_{[0,t_{uuv}]} \left(d _{uuv}\geq d_l \right) \right] \land \\
    &G \left[d_{uuv}>d_u \implies F_{[0,t_{uuv}]} \left(d_{uuv} \leq d_u \right) \right] \nonumber
\end{align}
where $d$ is the UUV's distance to the pipeline in the plane, $d_u$ and $d_l$ are the upper and lower distances to the pipe which the UUV must lie within and $t_{uuv}$ is the amount of time in which the UUV is allowed to leave this distance range.
Intuitively, this STL formula states that if the UUV leaves the ideal range of distances to the pipeline, $[d_l,d_u]$, then it must return to that ideal range within $t_{uuv}$ seconds. This ensures that the side scan sonar will be able to clearly track the pipeline while allowing the UUV to briefly lose the pipeline if needed, e.g. to avoid an obstacle. 

\subsubsection{Nominal System (Linearized Model)}
\label{sec:UUV_nominal_system}
Recall that we consider a high-fidelity UUV simulator based on the ROS-Gazebo real-time physics engine presented in \cite{Manhaes2016}. As mentioned in Section~\ref{sec:problem_statement}, executing the simulator requires significant computation resources, which makes it challenging to train a controller using RL. To overcome this issue, we use a linear model from the simulator and treat that model as the nominal system.

The linearized model of the UUV is available online at \url{https://github.com/Verisig/verisig}. In summary the linearization process is as follows. 
Using data from the real-time physics engine, one can identify a 4-th order linearized model of the UUV, which takes as input heading, speed, and depth commands and outputs the actual heading, speed, and depth of the UUV. Note that the full UUV model is still not linear since the linearized model does not capture global position. One can use a kinematic model, based on the speed and heading, in order to model the UUV position dynamics.


The equations in \eqref{eq:uuvsystemlin} show the dynamics of the linearized model of the UUV, where $f_{lin}$ represents the linearized dynamics. The model states are: $X = \begin{bmatrix} x_{uuv} & y_{uuv} & \theta_{uuv} &  v_{uuv} & d_{uuv} \end{bmatrix} ^{T}$, which contains the 2-D position, heading, speed and depth of the UUV, respectively. The measurements are the UUV's heading and distance with respect to the pipeline, i.e., $Y = \begin{bmatrix}\theta_{uuv} & d_{uuv} \end{bmatrix} ^{T}$, with the addition of additive Gaussian noise $W(t)$.
Finally, the NN controller takes the measurements as input and outputs a desired heading
\begin{subequations}\label{eq:uuvsystemlin}
\begin{align}
    X(t+1)&=f_{lin}(X(t),\text{NN}(Y(t))), \; X(0)=X_0\\
    Y(t)&=g(X(t),W(t)).
\end{align}
\end{subequations}

\subsubsection{Perturbed System (High-Fidelity Simulator)}
\label{sec:UUV_perturbed_system}

As mentioned above, we consider the high-fidelity UUV physics simulator to be the (real) perturbed system in this case study. In this case, the dynamics $f$ and state $\overline{X}(t)$ are internal to the real-time physics engine and thus can be considered unknown, as shown in \eqref{eq:uuvsystemphys}. Moreover, $\overline{Y}(t)$ and NN are the same as their counterparts in~\eqref{eq:uuvsystemlin}.
\begin{subequations}\label{eq:uuvsystemphys}
\begin{align}
    \overline{X}(t+1)&=f(\overline{X}(t),\text{NN}(\overline{Y}(t))), \; \overline{X}(0)=X_0\\
    \overline{Y}(t)&=g(\overline{X}(t)).
\end{align}
\end{subequations}





\subsubsection{Controller Training}
\label{sec:UUV_controller_training}

Next, we used the linearized system in \eqref{eq:uuvsystemlin} to train NN controllers to satisfy the objective in \eqref{eq:UUVspec} via RL. In particular, we  used the Twin Delayed Deep Deterministic Policy Gradients (TD3) algorithm \cite{Fujimoto2018} to train 4 NN controllers for guiding the UUV alongside a pipeline on the seafloor. Each NN consists of 2 tanh layers, each of width 32. The reward function $r$ is as follows:
\begin{align}
    r(X(t),\text{NN}(Y(t))) = 0.5 &- 100\; \mathbbm{1}_{d_{uuv} \not\in [d_1,d_2]} \\
    &- 0.5\; \text{NN}(Y(t)) - 0.1 \; \left|d_{uuv}-\frac{d_2-d_1}{2}\right| \nonumber
\end{align}
where $[d_1,d_2]$ is the safe interval which gets altered when training each of the NNs and $X(t)$, $Y(t)$, NN$(\cdot)$ and $d_{uuv}$ are as in Section \ref{sec:UUV_nominal_system}.


\subsubsection{UUV Experiments}
\label{sec:UUV_experiments}

To verify the risk of each of these NN controllers, each of them was used to generate 5000 runs of the linearized model and 300 runs of the real-time physics engine, each corresponding to 3 minutes of data per run.  Finally, the STL robustness risk of the STL specification in \eqref{eq:UUVspec}  is estimated as explained in Section ~\ref{sec:stoch} for the VaR and the CVaR and for a risk level of $\beta:=0.9$ and  a confidence of $1-\delta:=0.95$. The results are shown in Table \ref{Tab:UUVRisks}. The controllers which exhibit higher risk on the linearized model also exhibit higher risk on the real-time physics engine for both VaR and CVaR, supporting the risk verification gap theory from Section \ref{sec:risk_gap}. In particular, the safety range $10$ and $50$ controllers have the lowest risks on the linearized model and on the real-time physics engine, while the safety range $15$ controller has high risk both the linearized model and the real-time physics engine. This offers further empirical evidence that systems which exhibit lower robustness risk in simulation will also exhibit lower robustness risk in the real world.

\begin{table*}[h]
\caption{UUV risks for a real-time physics engine and a linearized model for specification \ref{eq:UUVspec}.}
    \centering
    \begin{tabular}{|c|c|c|c|c|}
    \hline \specialcell{Safety \\ Range} & \specialcell{Linearized \\ Model VaR} & \specialcell{Real-Time Physics \\ Engine VaR} & \specialcell{Linearized \\ Model CVaR} & \specialcell{Real-Time Physics \\ Engine CVaR} \\ \hline
    10 & \textbf{-10.1215} & -7.9380 & -7.0261 & -4.3691 \\ \hline
    15 & -1.8944 & 5.0788 & 2.3087 & 13.1208 \\ \hline
    30 & -8.2857 & -6.0316 & -5.2708 & 8.7808 \\ \hline
    50 & -10.0697 & \textbf{-8.3995} & \textbf{-7.3678} & \textbf{-5.4930} \\
    \hline
    \end{tabular}
    \label{Tab:UUVRisks}
\end{table*}

\begin{figure}
\centering
\includegraphics[scale=0.5]{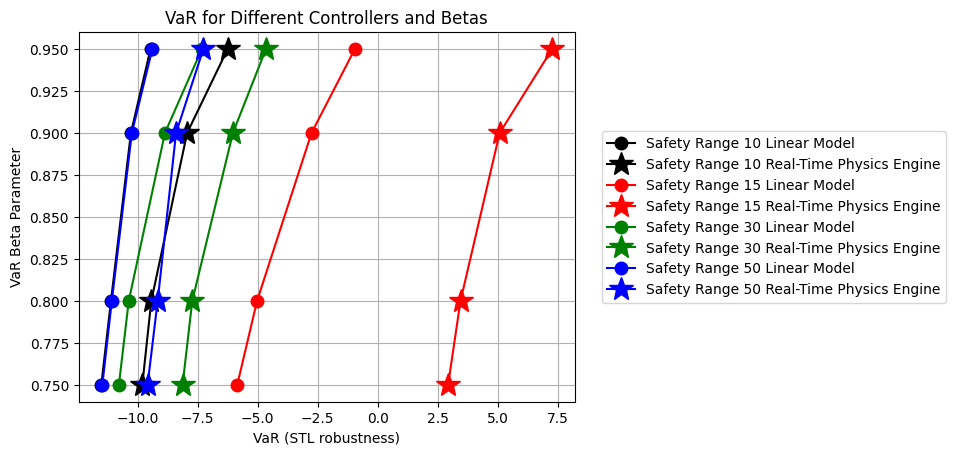}
\caption{Value-at-Risks (VaRs) for the STL robustness under the specification in \ref{eq:UUVspec} of the 4 different controllers on both UUV models for varied $\beta$ parameters.}
\label{fig:uuvVaRsManyBetas}
\end{figure}

\begin{figure}
\centering\includegraphics[scale=0.5]{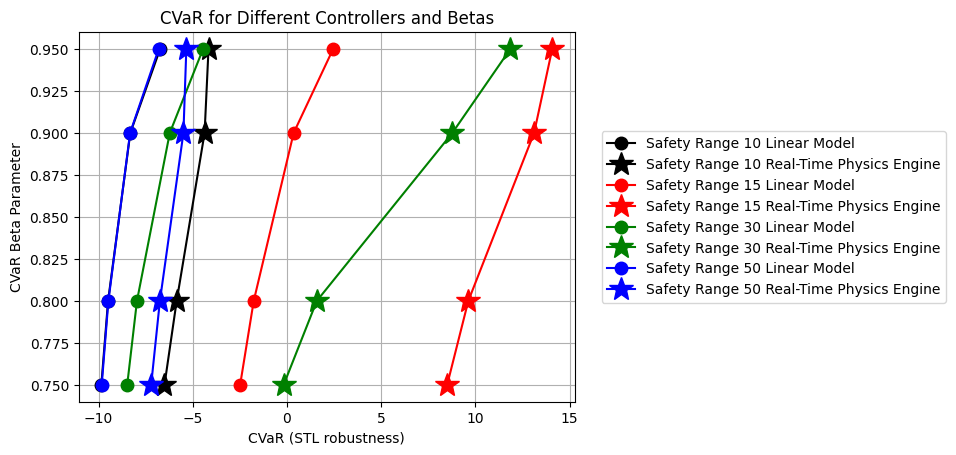}
\caption{Conditional Value-at-Risks (CVaRs) for the STL robustness under the specification in \ref{eq:UUVspec} of the 4 different controllers on both UUV models for varied $\beta$ parameters.}
\label{fig:uuvCVaRsManyBetas}
\end{figure}

Next, we varied the risk level $\beta$  for the VaR and CVaR metrics. These $\beta$ values control how much or little of the tail gets considered in the risk. Larger values only consider the extremes of the tails. Figures \ref{fig:uuvVaRsManyBetas} and \ref{fig:uuvCVaRsManyBetas} show the VaR and CVaR values, respectively, for both the linearized model and real-time physics engine and different controllers for the specification given in \ref{eq:UUVspec}. 
These figures both show that controllers which exhibit low risk values on the linearized model also exhibit low risk values on the real-time physics engine and vice versa, and are hence in line with the results in Section \ref{sec:risk_gap}.

\begin{figure}[t!] 
\begin{subfigure}[t]{0.6\textwidth}
\includegraphics[width=\linewidth]{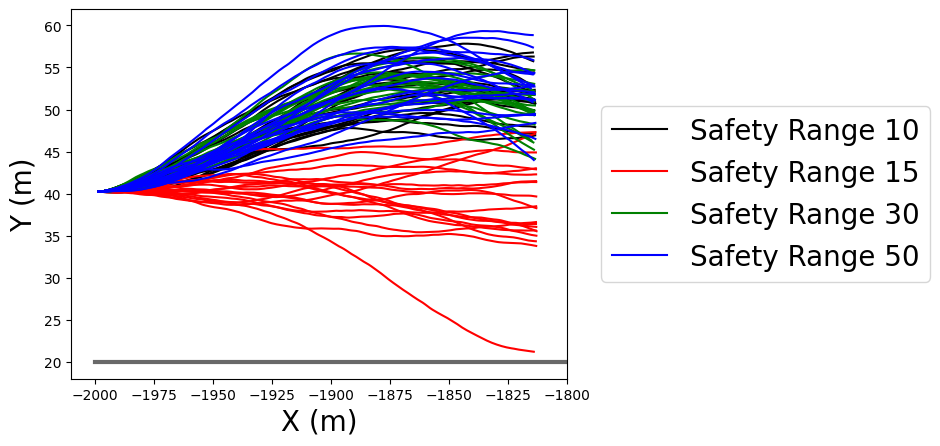}
\caption{UUV linear simulator traces} \label{fig:UUlinSimTraces}
\end{subfigure}\hspace*{\fill}
\begin{subfigure}[t]{0.39\textwidth}
\includegraphics[width=\linewidth]{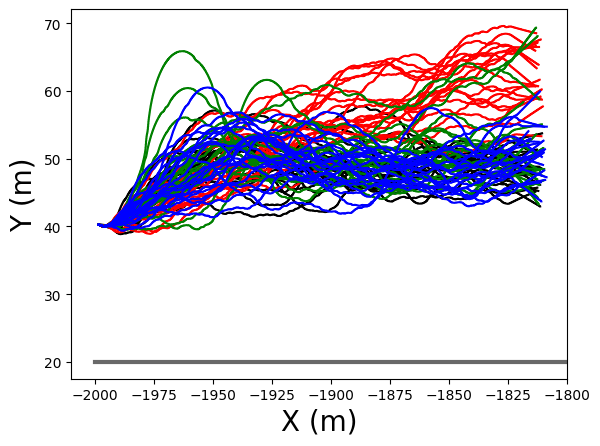}
\caption{UUV real-time physics engine traces} \label{fig:UUVPhysEngTraces}
\end{subfigure}
\caption{Traces for different UUV models and controllers.} \label{fig:UUVTraces}
\end{figure}

Finally we plot 20 traces at random for each controller for both the linear simulator and real-time physics engine, which can be seen in Figure \ref{fig:UUVTraces}. As expected, the safety range 15 controller paths get the farthest from the $[d_l,d_u]$ interval specified in the STL specification \ref{eq:UUVspec}, whereas the safety range 10 and 50 controllers stay most closely within the prescribed range. This matches the risks from Table \ref{Tab:UUVRisks}.

\subsubsection{Trace Difference Risk Bounds}
\label{sec:UUV_traceDiffBounds}

In this section, we use the results from Theorems \ref{thm:111_c} and \ref{thm:222_c} to bound the robustness risks of the real-time physics engine based on the robustness risks of the linearized model and the trace differences between the two models. We can define the trace difference bound $\Delta(\omega)$ as
\begin{align}\label{eq:traceDiffUUV}
     \Delta(\omega) := \sup_{t \in \mathbb{T}} \|\overline{X}(t,\omega)- X(t,\omega)\|
\end{align}
where $X$ denotes the linearized model and $\overline{X}$ denotes the real-time physics engine. The goal of these experiments is to experimentally determine bounds on \eqref{eq:traceDiffUUV} to use and verify Theorems \ref{thm:111_c} and \ref{thm:222_c}.

\begin{figure}[t!] 
\begin{subfigure}{0.48\textwidth}
\includegraphics[width=\linewidth]{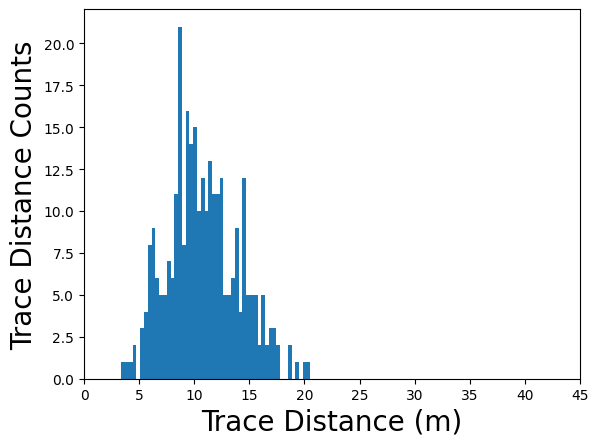}
\caption{Safety Range 10} \label{fig:UUVTraceDiffs10}
\end{subfigure}\hspace*{\fill}
\begin{subfigure}{0.48\textwidth}
\includegraphics[width=\linewidth]{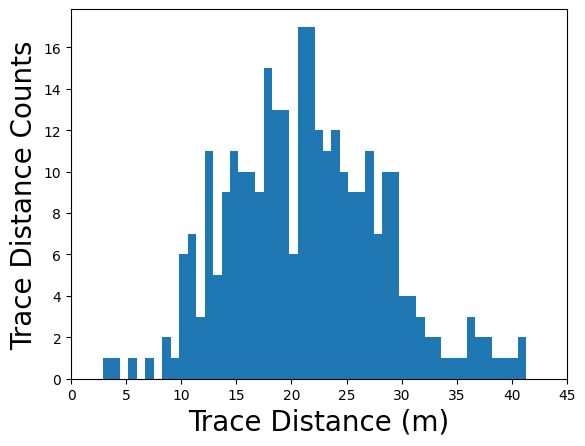}
\caption{Safety Range 15} \label{fig:f1-UUVTraceDiffs15}
\end{subfigure}

\medskip
\begin{subfigure}{0.48\textwidth}
\includegraphics[width=\linewidth]{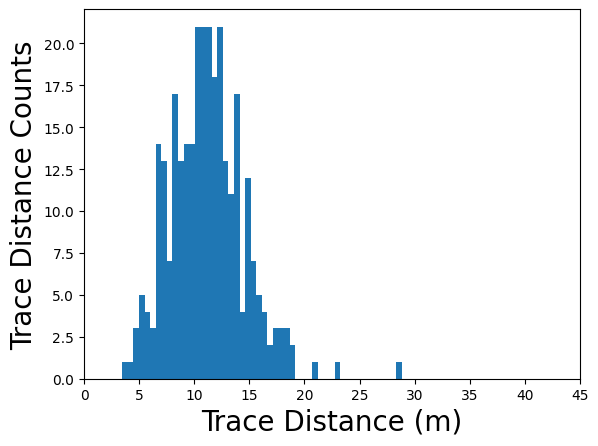}
\caption{Safety Range 30} \label{fig:f1-UUVTraceDiffs30}
\end{subfigure}\hspace*{\fill}
\begin{subfigure}{0.48\textwidth}
\includegraphics[width=\linewidth]{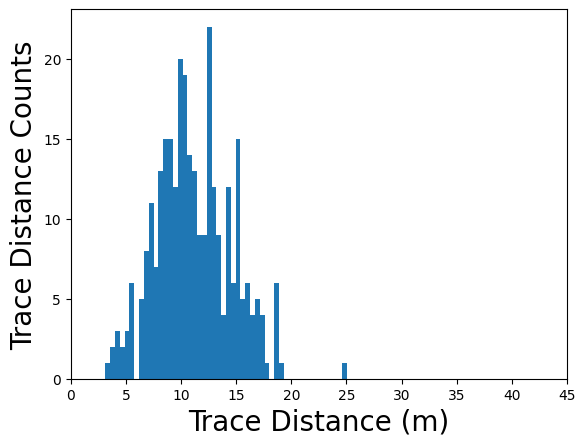}
\caption{Safety Range 50} \label{fig:f1-UUVTraceDiffs50}
\end{subfigure}

\caption{Trace Difference histograms between the linearized model and real-time physics engine for various controllers.} \label{fig:UUVTraceDiffs}
\end{figure}

To do this, we run both the linearized model and real-time physics engine from the same random initial conditions for $300$ trials to get a distribution of trace differences, which are shown as histograms in Figure \ref{fig:UUVTraceDiffs}. Then we compute the supremum, VaR, and CVaR of these trace differences, which we denote as $\sup_{\omega \in \Omega}\Delta(\omega)$, $VaR_\beta({\Delta(\omega)})$, and $CVaR_\beta(\Delta(\omega))$. We refer to these values as the trace difference risks. Finally, we use Theorems \ref{thm:111_c} and \ref{thm:222_c} to combine these trace difference risks with the linearized model robustness risks to determine upper bounds on the real-time physics engine robustness risks. These bounds, along with the trace difference risks, are shown in Tables \ref{Tab:UUVTraceDiffVaRTable} and \ref{Tab:UUVTraceDiffCVaRTable}. The real-time physics engine risk bounds are all larger than the empirical real-time physics engine risks we computed in Table \ref{Tab:UUVRisks}, which shows that the bounds are sound. In addition, the relative values of the real-time physics engine risk bounds match those of the empirical real-time physics engine risks. Specifically, the safety range $10$ and $50$ controllers offer the best robustness risk values and bounds for both Theorems \ref{thm:111_c} and \ref{thm:222_c}, while the safety range $15$ controllers have a much larger robustness risk value and bound than the other controllers do. This indicates that our risk bounds in Theorems \ref{thm:111_c} and \ref{thm:222_c} can aid in determining which controllers will best transfer from a nominal to a perturbed system. One drawback is that the bounds are a bit larger than the empirical robustness risks, which indicates that the bounds can be conservative.

\begin{table*}[h]
\caption{Real-time physics engine VaR bounds estimated from trace bounds and the linearized model VaR bounds.}
    \centering
    \begin{tabular}{|c|c|c|c|c|c|}
    \hline \specialcell{Safety \\ Range}   & \specialcell{Linearized\\ Model VaR} & \specialcell{$\sup_{\omega \in \Omega}(\Delta(\omega))$} & \specialcell{$VaR(\Delta(\omega))$} & \specialcell{Real-Time Physics \\ Engine VaR Bound \\ Theorem \ref{thm:111_c}}   & \specialcell{Real-Time Physics \\ Engine VaR Bound \\ Theorem \ref{thm:222_c}} \\ \hline
    10 & \textbf{-10.1215} & \textbf{20.5451} & \textbf{15.3278} & \textbf{10.4236} & \textbf{5.2063} \\ \hline
    15 & -1.8944 & 41.2857 & 29.9825 & 39.3913 & 28.0881 \\ \hline
    30 & -8.2857 & 28.8160 & 15.6049 & 20.5303 & 7.3192 \\ \hline
    50 & -10.0697 & 25.0174 & 15.7795 & 14.9477 & 5.7098 \\ 
    \hline
    \end{tabular}
    \label{Tab:UUVTraceDiffVaRTable}
\end{table*}

\begin{table*}[h]
\caption{Real-time physics engine CVaR bounds estimated from trace bounds and the linearized model CVaR bounds.}
    \centering
    \begin{tabular}{|c|c|c|c|c|c|}
    \hline \specialcell{Safety \\ Range}   & \specialcell{Linearized\\ Model VaR} & \specialcell{$\sup_{\omega \in \Omega}(\Delta(\omega))$} & \specialcell{$CVaR(\Delta(\omega))$} & \specialcell{Real-Time Physics \\ Engine CVaR Bound \\ Theorem \ref{thm:111_c}}   & \specialcell{Real-Time Physics \\ Engine CVaR Bound \\ Theorem \ref{thm:222_c}} \\ \hline
    10 & -7.0261 & \textbf{20.5451} & \textbf{20.0697} & \textbf{13.5190} & \textbf{13.0436} \\ \hline
    15 & 2.3087 & 41.2857 & 40.7459 & 43.5944 & 43.0546 \\ \hline
    30 & -5.2708 & 28.8160 & 26.7425 & 23.5452 & 21.4717 \\ \hline
    50 & \textbf{-7.3678} & 25.0174 & 23.5814 & 17.6496 & 16.2136 \\
    \hline
    \end{tabular}
    \label{Tab:UUVTraceDiffCVaRTable}
\end{table*}

%% file: chapter/conclusion.tex
\section{Conclusions}
\label{sec:conclusion}


In this paper, we have presented a framework for the closed-loop risk verification of stochastic systems with NN controllers. Specifically, our framework is based on risk metrics and distributions of trace robustness that are defined with respect to the violation of the system specification. We particularly allow for the use of formal specification languages such as signal temporal logic specifications. It was shown how such risk can be reliably estimated from finite data.  Additionally, we bound the changes in risk  when the system experiences disturbances for both general and iISS systems. We note that iISS systems have much tighter risk bounds. These bounds are conservative for the general system or require the strong iISS property, so we also provide an empirical method to determine the change in risk based on differences of sampled traces. Finally, our framework was applied to the risk verification and empirical risk gap estimation of an F1/10 race car and an underwater vehicle controlled by end-to-end NN controllers.
Future research directions include applying the risk verification framework to NNs in an open loop or standalone setting, using the framework to guide the learning process of NN controllers, and extending the risk gap quantification to the sim2real gap.


We believe that this work makes three important contributions for the analysis of  artificial intelligent systems. First, risk estimation is an essential tool for the verification of safety-critical systems with NN components. Notably, our method can directly verify closed-loop stochastic systems with NN controllers, whereas current model-based verification techniques cannot. Second, NN controllers are  complex and their closed-loop verification imposes significant computational costs, whereas our data-driven method trades off this computational complexity with the ability to collect data. Lastly, the quantification of the risk verification gap allows us to quantify the worst case risk of a perturbed version of a nominal system. We further believe that the information gained from risk verification may in the future guide the design process and close the loop between training and verification of NN controllers.

%% file: chapter/appendix.tex
\section{Risk Metrics}
\label{app:risk}
We next summarize some desirable properties of a risk metric. Let therefore $Z,Z'\in \mathfrak{F}(\Omega,\mathbb{R})$ be cost random variables. A risk metric is  \emph{coherent} if the following four properties are satisfied.\\
\emph{1. Monotonicity:} If $Z(\omega) \leq Z'(\omega)$ for all $\omega\in\Omega$, it holds that $R(Z) \le R(Z')$.\\
\emph{2. Translation Invariance:} Let $c\in\mathbb{R}$. It holds that $R(Z + c) = R(Z) + c$.\\
\emph{3. Positive Homogeneity:} Let $c\in\mathbb{R}_{\ge 0}$. It holds that $R(c Z) =  R(Z)$.\\
\emph{4. Subadditivity:} It holds that $R(Z + Z') \leq R(Z) + R(Z')$.

If the risk metric additionally satisfies the following two properties, then it is called a distortion risk metric. 

\noindent \emph{5. Comonotone Additivity:} If $(Z(\omega) - Z(\omega'))(Z'(\omega) - Z'(\omega')) \ge 0$ for all $\omega, \omega' \in \Omega$ (namely, $Z$ and $Z'$ are commotone), it holds that $R(Z + Z') = R(Z) + R(Z')$.\\
\emph{6. Law Invariance:} If $Z$ and $Z'$ are identically distributed, then $R(Z) = R(Z')$.

Common examples of popular risk metrics are the expected value $\text{E}(Z)$ (risk neutral) and the worst-case $\text{ess} \sup_{\omega\in\Omega} Z(\omega)$ as well as:
\begin{itemize}
	\item Mean-Variance: $\text{E}(Z) + \lambda \text{Var}(Z)$ where  $c> 0$.
	\item Value at Risk (VaR) at level $\beta \in (0,1)$: $VaR_\beta(Z):=\inf \{ \alpha \in \mathbb{R} |  F_Z(\alpha) \ge \beta \}$.
	\item Conditional Value at Risk (CVaR) at level $\beta \in (0,1)$: $CVaR_\beta(Z):=E(Z|Z>VaR_\beta(Z))$.
\end{itemize}

Many risk metrics are not coherent and can lead to a misjudgement of risk, e.g., the mean-variance is not monotone and the  value at risk (which is closely related to chance constraints as often used in optimization) does not satisfy the subadditivity property.

\section{Measurability of $-\rho^c(X)$}
\label{app:measurability}
For a  constraint function $c:\mathbb{R}^n\times \mathbb{T}\to\mathbb{R}$, we want to show that the robustness function $\rho^c(X(\cdot,\omega))$ is measureable in $\omega$. In this case, $\rho^c(X)$ and hence $-\rho^c(X)$ are random variables so that $R(-\rho^c(X))$ is well defined. Recall that $\rho^c(X(\cdot,\omega))=\inf_{t\in\mathbb{T}} \, \text{Dist}^c(X(t,\omega),t)$ where the signed distance is
\begin{align*}
    \text{Dist}^c(X(t,\omega),t)=\begin{cases}
    \inf_{x'\in \text{cl}(O^{\neg c}(t))} \|X(t,\omega)-x'\| &\text{if } c(X(t,\omega),t)\ge 0\\
    -\inf_{x'\in \text{cl}(O^c(t))} \|X(t,\omega)-x'\| &\text{otherwise. }
    \end{cases}
\end{align*} 
We first show that $\text{Dist}^c(X(t,\omega))$ is measurable in $\omega$ for a fixed $t\in\mathbb{T}$. Note therefore that: 1) the function $c(X(t,\omega))$ is measurable in $\omega$ for a fixed $t\in\mathbb{T}$, 2) the indicator function of a measurable set is measurable, and 3) the distance function to a non-empty set, i.e., $\inf_{x'\in \text{cl}(O^{\neg c}(t))} \|X(t,\omega)-x'\|$, is continuous (see e.g., \cite[Chapter 3]{munkres1975prentice}) and hence measurable. Consequently, $\text{Dist}^c(X(t,\omega))$ is measurable. To now conclude measurability of $\rho^c(X(\cdot,\omega))$ in $\omega$, note that the infimum over a countable number of measurable functions is again measurable, see e.g., \cite[Theorem 4.27]{guide2006infinite}. 

\section{Proof of Lemma \ref{lemma:111_c}}
\label{app:lemma111_c}
For a  constraint function $c:\mathbb{R}^n\times \mathbb{T}\to\mathbb{R}$, we want to show that 
  $\rho^c(X(\cdot,\omega))-\Delta\le \rho^c(\overline{X}(\cdot,\omega))$
for each realization $\omega\in\Omega$. For convenience, let $\overline{x}:=\overline{X}(\cdot,\omega)$ and $x:=X(\cdot,\omega)$ and recall that the robustness function $\rho^c(x)$ for the constraint function $c$ is defined as $\rho^c(x)=\inf_{t\in\mathbb{T}} \, \text{Dist}^c(x(t),t)$
where the signed distance is
\begin{align*}
    \text{Dist}^c(x(t),t)=\begin{cases}
    \inf_{x'\in \text{cl}(O^{\neg c}(t))} \|x(t)-x'\| &\text{if } c(x(t),t)\ge 0\\
    -\inf_{x'\in \text{cl}(O^c(t))} \|x(t)-x'\| &\text{otherwise. }
    \end{cases}
\end{align*}

Note next that it holds that $\inf_{x'\in \text{cl}(O^{\neg c}(t))} \|x(t)-x'\|$ and $\inf_{x'\in \text{cl}(O^c(t))} \|x(t)-x'\|$ are Lipschitz continuous functions with Lipschitz constant one, see for instance \cite[Chapter 3]{munkres1975prentice}. Consequently, it holds that $\text{Dist}^c(x(t),t)$ is Lipschitz continuous in its first argument with Lipschitz constant one. From this, we have that  $|\text{Dist}^c(x(t),t)-\text{Dist}^c(\overline{x}(t),t)| \le \|x(t)-\overline{x}(t)\|\le \Delta$ for all times $t\in\mathbb{T}$. By the definition of $\rho^c$, we consequently have that $|\rho^c(x)-\rho^c(\overline{x})|=|\inf_{t\in\mathbb{T}}\text{Dist}^c(x(t),t)-\inf_{t\in\mathbb{T}}\text{Dist}^c(\overline{x}(t),t)| \le \Delta$ so that $\rho^c(\overline{x}) \ge \rho^c(x) - \Delta$.

\section{Proof of Theorem \ref{thm:111_c}}
\label{app:thm111_c}

For a  constraint function $c:\mathbb{R}^n\times \mathbb{T}\to\mathbb{R}$, we want to show that $R(-\rho^c(\overline{X}))\le R(-\rho^c(X))+\Delta$. By Lemma \ref{lemma:111_c} and since the risk metric $R$ is monotone  (see Appendix \ref{app:risk}), it follows that $R(-\rho^c(\overline{X}))\le R(-\rho^c(X)+\Delta)$. Since $R$ is additionally translationally invariant (see Appendix \ref{app:risk}), it further holds that $R(-\rho^c(X)+\Delta)=R(-\rho^c(X))+\Delta$ so that $R(-\rho^c(\overline{X}))\le R(-\rho^c(X))+\Delta$, which concludes the proof.

 \section{Proof of Theorem \ref{thm:222_c}}
\label{app:thm222_c}

For a  constraint function $c:\mathbb{R}^n\times \mathbb{T}\to\mathbb{R}$, we want to show that $R(-\rho^c(\overline{X}))\le R(-\rho^c(X))+R(\Gamma)$. By minor modification of Lemma \ref{lemma:111_c}, one can show that $ \rho^c(\overline{X}(\cdot,\omega))\ge \rho^c(X(\cdot,\omega)) - \Gamma(\omega)$ for each realization $\omega\in\Omega$. Since the risk metric $R$ is monotone (see Appendix \ref{app:risk}), it follows that $R(-\rho^c(\overline{X}))\le R(-\rho^c(X)+\Gamma)$. Since $R$ is now assumed to be subadditive (see Appendix \ref{app:risk}), it further holds that $R(-\rho^c(X)+\Gamma)\le R(-\rho^c(X))+R(\Gamma)$ so that $R(-\rho^c(\overline{X}))\le R(-\rho^c(X))+R(\Gamma)$, which concludes the proof.

\section{Bound $\Delta(t)$ on $\|\overline{X}(t,\omega)- X(t,\omega)\|$}
\label{app:riskGap}

First, recall that the nominal system has dynamics
\begin{align*}
    X(t+1)&=f(X(t),u(Y(t)),V(t)), \; X(0)=X_0\\
    Y(t)&=g(X(t),W(t)),
\end{align*}
while the perturbed system has dynamics 
\begin{align*}
    \overline{X}(t+1)&=f(\overline{X}(t),u(\overline{Y}(t)),\overline{V}(t)), \; \overline{X}(0)=X_0\\
    \overline{Y}(t)&=g(\overline{X}(t),\overline{W}(t)).
\end{align*}
Additionally, recall that $L_{f,1}$, $L_{f,2}$, and $L_{f,3}$ are the Lipschitz constants of the first, second, and third arguments of $f$, respectively, $L_{g,1}$ and $L_{g,2}$ are the Lipschitz constants of $g$, and $L_{u}$ is the Lipschitz constant of $u$. Finally, note that we have disturbance bounds $v^*:=2\max_{v\in D_v}\|v\|$ and $w^*:=2\max_{w\in D_w}\|w\|$. Note that initially $\| \overline{X}(0,\omega) - X(0,\omega) \|=0$ as $\overline{X}(0,\omega)=X_0(\omega)$ and ${X}(0,\omega)=X_0(\omega)$. We can then upper bound $\| \overline{X}(t+1,\omega) - X(t+1,\omega) \|$ recursively as follows.

\begin{align*}
    \| \overline{X}(t+1,\omega) - X(t+1,\omega) \|& \stackrel{(a)}{=} \| f(\overline{X}(t,\omega),u\left(\overline{Y}(t) \right),\overline{V}(t)) - f(X(t,\omega),u(Y(t)),V(t)) \| \\
    &\stackrel{(b)}{=} \| f(\overline{X}(t,\omega),u\left(g(\overline{X}(t,\omega),\overline{W}(t))\right),\overline{V}(t)) \\
    &\hspace{0.5cm}- f(X(t,\omega),u\left (g(X(t,\omega),W(t))\right),V(t)) \| \\
    & \stackrel{(c)}{\leq}  L_{f,1} \| \overline{X}(t,\omega) - X(t,\omega) \|  + L_{f,2} \| u\left(g(\overline{X}(t,\omega),\overline{W}(t))\right)  \\
    &\hspace{0.5cm} - u(g(X(t,\omega),W(t))) \| + L_{f,3}v^* \\
    & \stackrel{(d)}{\leq} L_{f,1} \Delta(t) + L_{f,2} L_{u} \|  g(\overline{X}(t,\omega),\overline{W}(t)) \\
    &\hspace{0.5cm} - g(X(t,\omega),W(t)) \| + L_{f,3}v^*\\
    & \stackrel{(e)}{\leq} L_{f,1} \Delta(t) + L_{f,2} L_{u} (L_{g,1} \| \overline{X}(t,\omega) - X(t,\omega) \| + L_{g,2}w^*) + L_{f,3}v^* \\
    & \stackrel{(f)}{\leq} L_{f,1} \Delta(t) + L_{f,2} L_{u} (L_{g,1} \Delta(t) + L_{g,2}w^*) + L_{f,3}v^*=\Delta(t+1). 
\end{align*}
Equalities $(a)$ and $(b)$ follow from plugging in the dynamics of the nominal and perturbed systems. Inequality $(c)$ follows by plugging in the Lipschitz constants of $f$ and applying the bounds on $V$ and $\overline{V}$. Inequality $(d)$ follows from plugging in the Lipschitz constant of $u$, while ineqaulity $(e)$ follows by the Lipschitz constants of $g$ and the bounds on $W$ and $\overline{W}$. Finally, inequality $(f)$ follows from plugging in the bound from the previous time step $\Delta(t)$.

\section{Semantics of STL}
\label{app:STL_sem}

We now define the semantics of an STL formula $\phi$ as in \eqref{eq:full_STL} for a signal $x$ and thereby give $\phi$ a meaning. To do so, we define the satisfaction function $\beta^\phi:\mathfrak{F}(T,\mathbb{R}^n)\times T \to \mathbb{B}$ where $\beta^\phi(x,t)=\top$ indicates that the signal $x$ satisfies the formula $\phi$ at time $t$, while $\beta^\phi(x,t)=\bot$ indicates that $x$ does not satisfy $\phi$ at time $t$. For a signal $x:\mathbb{N}\to\mathbb{R}^n$, the semantics $\beta^\phi(x,t)$ of an STL formula $\phi$ are inductively defined as
	\begin{align*}
	\beta^\top(x,t)&:=\top,  \\
	\beta^\mu(x,t)&:=\begin{cases}
	\top &\text{ if }	x(t)\in O^\mu\\	
	\bot &\text{ otherwise, }	
	\end{cases}\\
	\beta^{\neg\phi}(x,t)&:= \neg \beta^{\phi}(x,t),\\
	\beta^{\phi' \wedge \phi''}(x,t)&:=\min(\beta^{\phi'}(x,t),\beta^{\phi''}(x,t)),\\
	\beta^{\phi' U_I \phi''}(x,t)&:=\sup_{t''\in (t\oplus I)\cap \mathbb{N}}\Big( \min\big(\beta^{\phi''}(x,t''),\inf_{t'\in(t,t'')\cap \mathbb{N}}\beta^{\phi'}(x,t')\big)\Big).\\
	\end{align*}
In the above definition, the symbols $\oplus$ and $\ominus$ denote the Minkowski sum and the Minkowski difference, respectively. Finally, an STL formula $\phi$ is said to be satisfiable if $\exists x\in \mathfrak{F}(T,\mathbb{R}^n)$ such that $\beta^\phi(x,0)=\top$.

We say that an STL formula $\phi$ is bounded if all time intervals $I$ in $\phi$ are bounded. A bounded STL formula $\phi$ has a formula length $L^\phi$ that can be calculated, similarly to \cite{sadraddini2015robust}, as
 \begin{align*}
     L^\top&=L^\mu:=0\\
     L^{\neg\phi}&:=L^\phi\\
     L^{\phi'\wedge\phi''}&:=\max(L^{\phi'},L^{\phi''})\\
     L^{\phi' U_I \phi''}&:=\max \{I\cap \mathbb{N}\}+\max(L^{\phi'},L^{\phi''}).
 \end{align*}

A finite signal of length $L^\phi$ is now sufficient to determine if $\phi$ is satisfied at time $t$. In particular, a signal $x:T_L\to\mathbb{R}^n$, where $T_L:=\{t,\hdots,t,\hdots,t+L^\phi\}$, contains sufficient information to determine if $\phi$ is satisfied by the signal $x$ at time $t$.

\section{Robustness Degree and Robust Semantics of STL}
\label{app:STL}

Let us define the set of  signals that satisfy $\phi$ at time $t$ as
	\begin{align*}
	\mathcal{L}^\phi(t):=\{x\in \mathfrak{F}(T,\mathbb{R}^n)|\beta^\phi(x,t)=\top\}.
	\end{align*} 
	For a signal $x\in \mathcal{L}^\phi(t)$, the robustness degree $\text{dist}^\phi(x,t)$ then tells us how much the signal $x$ can be perturbed by additive noise before changing the Boolean truth value of the specification $\phi$. Towards a formal definition, let us define the signal metric
	\begin{align*}
	\kappa(x,x^*):=\sup_{t\in T} \|x(t)-x^*(t)\|.
	\end{align*}
	Note that $\kappa(x,x^*)$ is the $L_\infty$ norm of the signal $x-x^*$. The distance of $x$ to the set $\mathcal{L}^\phi(t)$ is then defined via the metric $\kappa$ as 
	\begin{align*}
	\text{dist}^\phi(x,t)=\bar{\kappa}\big(x,\text{cl}(\mathcal{L}^\phi(t))\big):=\inf_{x^*\in \text{cl}(\mathcal{L}^\phi(t))}\kappa(x,x^*),
	\end{align*} 
	where $\text{cl}(\mathcal{L}^\phi(t))$ denotes the closure of $\mathcal{L}^\phi(t)$.

As remarked the function $\text{dist}^\phi(x,t)$ is hard to evaluate. Instead, the robust semantics that are associated with an STL formula $\phi$ can be used. For a signal $x:\mathbb{N}\to\mathbb{R}^n$, the robust semantics $\rho^\phi(x,t)$  of an STL formula $\phi$ are inductively defined as 
	\begin{align*}
	\rho^{\top}(x,t)& := \top,\\
	\rho^{\mu}(x,t)& := \begin{cases} \text{dist}^{\neg\mu}(x,t) &\text{if } x\in \mathcal{L}^\mu(t)\\
	-\text{dist}^{\mu}(x,t) &\text{otherwise,}
	\end{cases}\\
	\rho^{\neg\phi}(x,t) &:= 	-\rho^{\phi}(x,t),\\
	\rho^{\phi' \wedge \phi''}(x,t) &:= 	\min(\rho^{\phi'}(x,t),\rho^{\phi''}(x,t)),\\
	\rho^{\phi' U_I \phi''}(x,t) &:= \underset{t''\in (t\oplus I)\cap \mathbb{N}}{\text{sup}}  \Big(\min\big(\rho^{\phi''}(x,t''),\underset{t'\in (t,t'')\cap \mathbb{N}}{\text{inf}}\rho^{\phi'}(x,t') \big)\Big), \\
	\end{align*}
Note that the difference between the non-robust and robust STL semantics lies solely in the way these operators are defined for predicates $\mu$.

\section{Proof of Lemma \ref{lemma:1}}
\label{app:lemma1}

Our goal is to show that the inequalities in \eqref{eq:phi_inequ} and \eqref{eq:phi_inequ_} hold. We present the proof for \eqref{eq:phi_inequ_} first, i.e., the case of a bounded STL formula $\phi$ with formula length $L^\phi$ and for the trajectory error $\Delta(t)$ as in \eqref{eq_Lip}. The result for \eqref{eq:phi_inequ} then follows the same way with minor modifications and is omitted. For an STL formula $\phi$ as in \eqref{eq:full_STL}, we hence want to show that $\rho^\phi(X(\cdot,\omega),t)-\Delta(t+L^\phi)\le \rho^\phi(\overline{X}(\cdot,\omega),t))$ for each realization $\omega\in\Omega$ and time $t\in\mathbb{T}$.

For convenience, let again $\overline{x}:=\overline{X}(\cdot,\omega)$ and $x:=X(\cdot,\omega)$. The proof follows inductively on the structure of $\phi$ as in \eqref{eq:full_STL} and without considering the case of negations as $\phi$ is assumed to be in positive normal form. 

\emph{Predicates. } First recall the STL robust semantics for predicates defined in Appendix \ref{app:STL} as
\begin{align*}
    \rho^{\mu}(x,t)& := \begin{cases} \text{dist}^{\neg\mu}(x,t) &\text{if } x\in \mathcal{L}^\mu(t)\\
	-\text{dist}^{\mu}(x,t) &\text{otherwise,}
	\end{cases}
\end{align*}
and note that
\begin{align*}
	\text{dist}^{\mu}(x,t)= \bar{d}\big(x(t),\text{cl}(O^\mu)\big):=\inf_{x'\in \text{cl}(O^\mu)} \|x(t)-x'\|.
	\end{align*} 
	according to \cite[Lemma 57]{fainekos2009robustness}. It further holds that $\bar{d}$ is a Lipschitz continuous function with Lipschitz constant one, see for instance \cite[Chapter 3]{munkres1975prentice}. Accordingly, it holds that $\rho^{\mu}(x,t)$ is Lipschitz continuous in its first argument with Lipschitz constant one. For each predicate $\mu_i$ and for each time $t'\le t+L^\phi$, it hence follows that  $|\rho^{\mu_i}(x,t')-\rho^{\mu_i}(\overline{x},t')| \le \|x(t')-\overline{x}(t')\|\le \Delta(t')\le \Delta(t+L^\phi)$ where the last inequality follows by  since $\Delta(t')\le \Delta(t+L^\phi)$ for $t'\le t+L^\phi$. Consequently, we have that $ \rho^{\mu_i}(\overline{x},t')\ge \rho^{\mu_i}(x,t)- \Delta(t+L^\phi)$ for $t'\le t+L^\phi$.

\emph{Conjunctions. } For conjunctions $\phi'\wedge\phi''$ and by the induction assumption, it holds that $ \rho^{\phi'}(\overline{x},t')\ge \rho^{\phi'}(x,t')- \Delta(t+L^\phi)$ and $ \rho^{\phi''}(\overline{x},t')\ge \rho^{\phi''}(x,t')- \Delta(t+L^\phi)$ for $t'\le t+L^{\phi}-L^{\phi'\wedge\phi''}$. Now, for $t'\le t+L^{\phi}-L^{\phi'\wedge\phi''}$, it follows that
\begin{align*}
    \rho^{\phi'\wedge \phi''}(\overline{x},t')&=\min(\rho^{\phi'}(\overline{x},t'),\rho^{\phi'}(\overline{x},t'))\\
    &\ge \min(\rho^{\phi'}(x,t'),\rho^{\phi''}(x,t'))- \Delta(t+L^\phi)\\
    &=\rho^{\phi'\wedge \phi''}(x,t')- \Delta(t+L^\phi).
\end{align*}

\emph{Until. } Using the induction assumption and the same reasoning as for conjunctions, it follows that $ \rho^{\phi'U_I \phi''}(\overline{x},t')\ge \rho^{\phi'U_I \phi''}(x,t')- \Delta(t+L^\phi) \; $ for $t'\le t+L^\phi- L^{\phi'U_I \phi''}$, and it can hence be concluded that \eqref{eq:phi_inequ} holds.

\section{Proof of Theorem \ref{thm:1}}
\label{app:thm1}
 We  present the proof for a bounded STL formula $\phi$ with formula length $L^\phi$ and for the trajectory error $\Delta(t)$ as in \eqref{eq_Lip} first, i.e., we first show that $R(-\rho^\phi(\overline{X},t))\le R(-\rho^\phi(X,t))+\Delta(t+L^\phi)$. The proof for $R(-\rho^\phi(\overline{X},t))\le R(-\rho^\phi(X,t))+\Delta$ then follows the same way, i.e., for unbounded STL formulas $\phi$ and for the trajectory error $\Delta$. 
 By \eqref{eq:phi_inequ} and since the risk metric $R$ is monotone (see Appendix \ref{app:risk}), it follows that $R(-\rho^\phi(\overline{X},t))\le R(-\rho^\phi(X,t)+\Delta(t+L^\phi))$. Since $R$ is additionally translationally invariant (see Appendix \ref{app:risk}), it further holds that $R(-\rho^\phi(X,t)+\Delta(t+L^\phi))=R(-\rho^\phi(X,t))+\Delta(t+L^\phi)$ so that $R(-\rho^\phi(\overline{X},t))\le R(-\rho^\phi(X,t))+\Delta(t+L^\phi)$, which concludes the proof.

\section{F1/10 GAN Figures}
\label{app:GANFigures}
\begin{figure}[t!] 
\begin{subfigure}{0.3\linewidth}
\includegraphics[width=\linewidth]{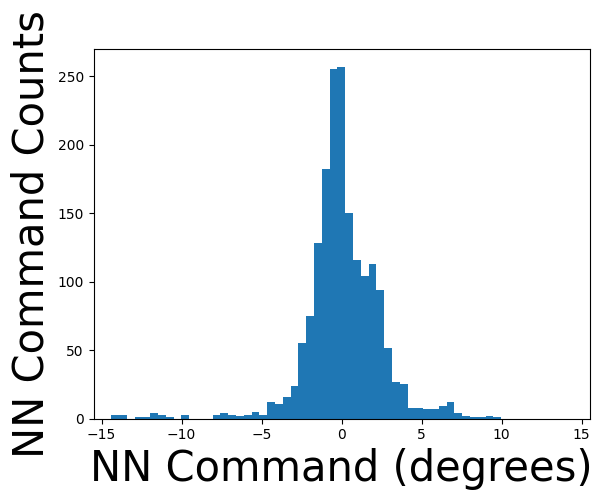}
\caption{Controller 1 Missing Ray} \label{fig:f1-10MissingRayCommandsC1}
\end{subfigure}
\begin{subfigure}{0.3\linewidth}
\includegraphics[width=\linewidth]{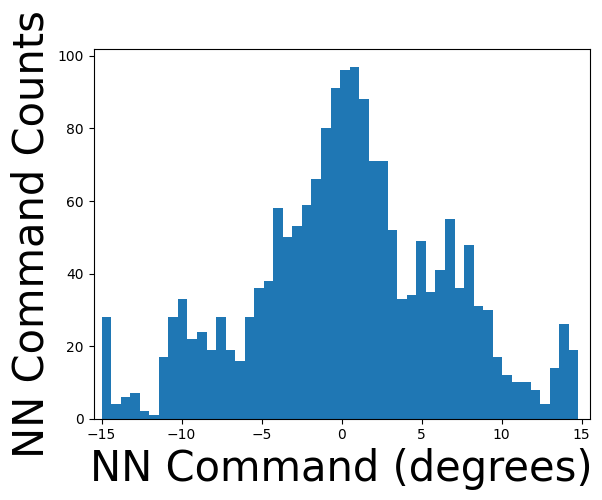}
\caption{Controller 1 GAN} \label{fig:f1-10ganCommandsC1}
\end{subfigure}
\begin{subfigure}{0.3\linewidth}
\includegraphics[width=\linewidth]{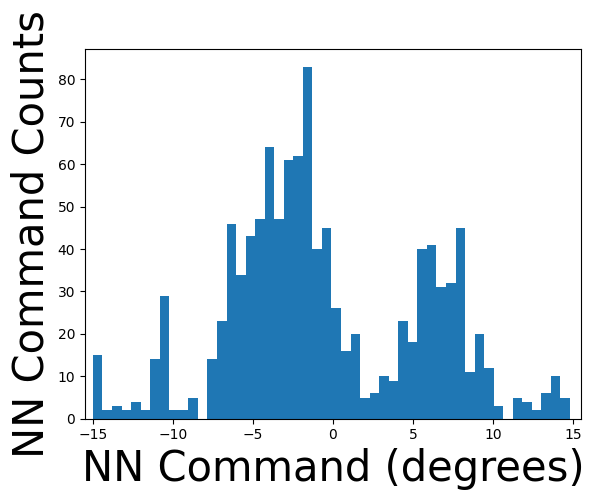}
\caption{Controller 1 Real LiDAR} \label{fig:f1-10realLiDARCommandsC1}
\end{subfigure}
\begin{subfigure}{0.32\textwidth}
\includegraphics[width=\linewidth]{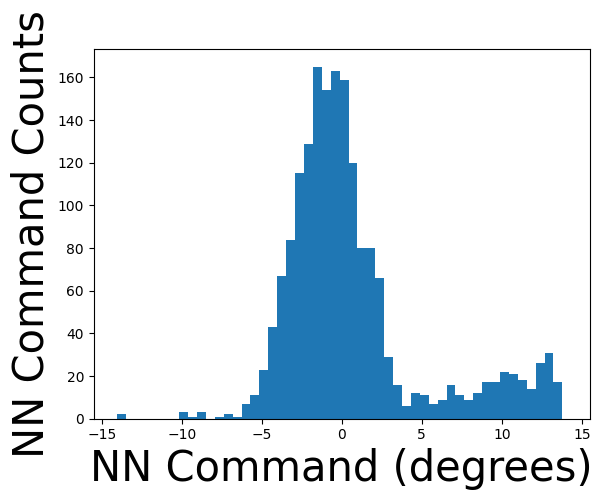}
\caption{Controller 2 Missing Ray} \label{fig:f1-10MissingRayCommandsC2}
\end{subfigure}
\begin{subfigure}{0.32\textwidth}
\includegraphics[width=\linewidth]{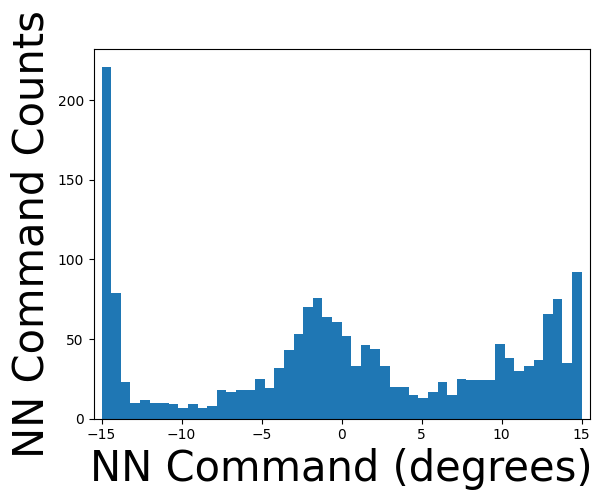}
\caption{Controller 2 GAN} \label{fig:f1-10ganCommandsC2}
\end{subfigure}\hspace*{\fill}
\begin{subfigure}{0.32\textwidth}
\includegraphics[width=\linewidth]{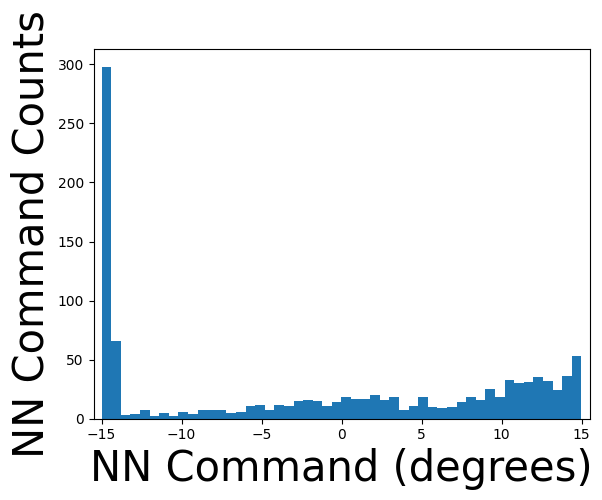}
\caption{Controller 2 Real LiDAR} \label{fig:f1-10realLiDARCommandsC2}
\end{subfigure}
\begin{subfigure}{0.32\textwidth}
\includegraphics[width=\linewidth]{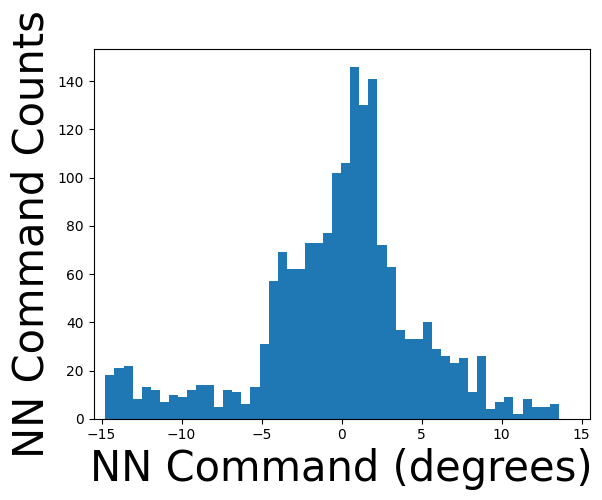}
\caption{Controller 3 Missing Ray} \label{fig:f1-10MissingRayCommandsC3}
\end{subfigure}
\begin{subfigure}{0.32\textwidth}
\includegraphics[width=\linewidth]{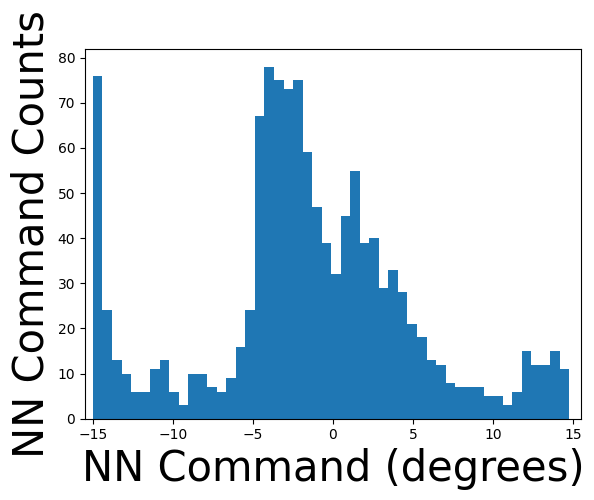}
\caption{Controller 3 GAN} \label{fig:f1-10ganCommandsC3}
\end{subfigure}\hspace*{\fill}
\begin{subfigure}{0.32\textwidth}
\includegraphics[width=\linewidth]{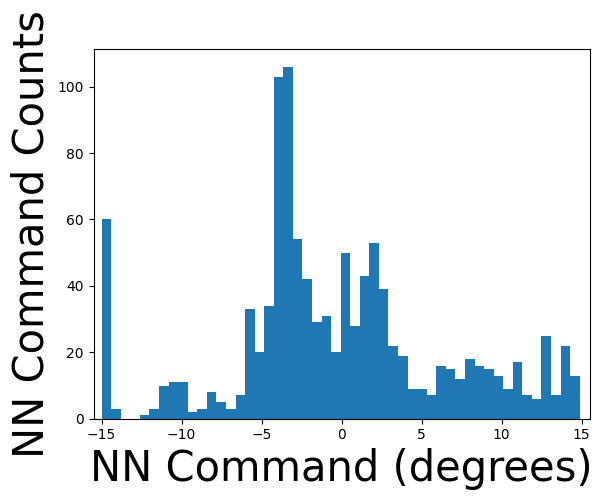}
\caption{Controller 3 Real LiDAR} \label{fig:f1-10realLiDARCommandsC3}
\end{subfigure}
\begin{subfigure}{0.32\textwidth}
\includegraphics[width=\linewidth]{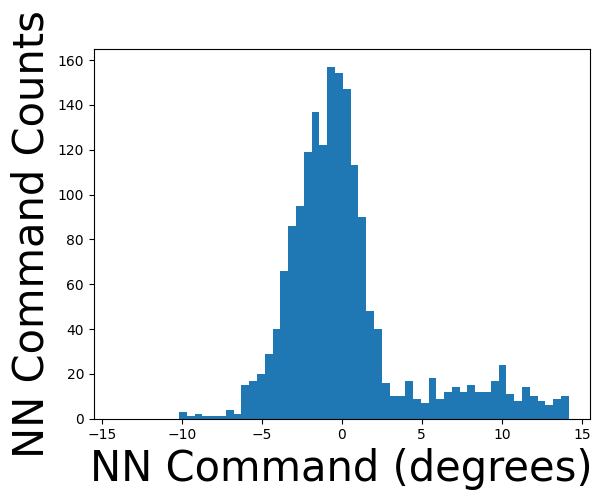}
\caption{Controller 6 Missing Ray} \label{fig:f1-10MissingRayCommandsC6}
\end{subfigure}
\begin{subfigure}{0.32\textwidth}
\includegraphics[width=\linewidth]{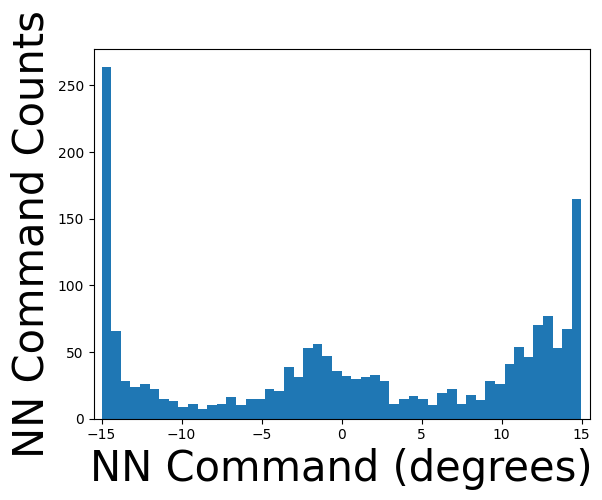}
\caption{Controller 6 GAN} \label{fig:f1-10ganCommandsC6}
\end{subfigure}\hspace*{\fill}
\begin{subfigure}{0.32\textwidth}
\includegraphics[width=\linewidth]{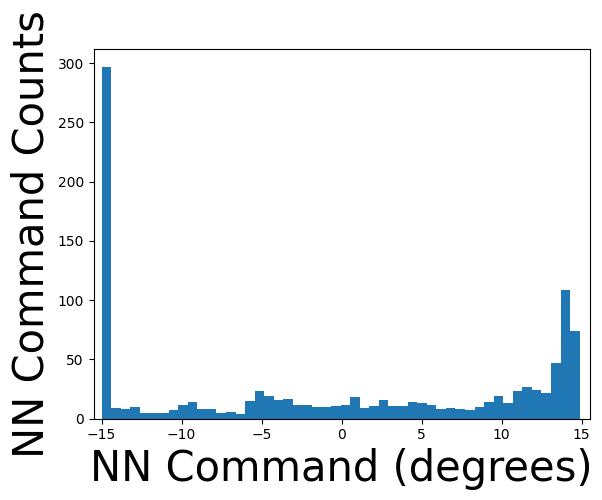}
\caption{Controller 6 Real LiDAR} \label{fig:f1-10realLiDARCommandsC6}
\end{subfigure}

\caption{Histograms showing control command distributions for the GAN LiDAR model and real LiDAR data for various controllers.} \label{fig:f1-10GanRealLiDARHists_2}
\end{figure}

In this section, we report in Figure \ref{fig:f1-10GanRealLiDARHists_2} the remaining histograms for the GAN evaluation from Section \ref{sec:ganAnalysis}.

\clearpage